\newif\ifextended 
\newif\ifccblock 
\newif\ifauthor 
\newif\ifsmall 
\newif\iforcid 
\newif\ifcolor 
\newif\ifthesis 
\newif\iftprint 

\extendedtrue\ccblockfalse\authortrue\smallfalse\orcidtrue\colortrue\thesisfalse\tprintfalse

\documentclass[runningheads]{llncs}

\ifsmall
  \usepackage[paperheight=235mm,paperwidth=155mm,textwidth=12.2cm,textheight=19.3cm]{geometry}
\fi

\newcommand{\gfivewidth}{169mm}
\newcommand{\gfiveheight}{239mm}
\newlength{\evenmargin}
\ifthesis
  \usepackage[paperwidth=\gfivewidth, paperheight=\gfiveheight, textwidth=\textwidth, textheight=\textheight \iftprint\else, inner=\evenmargin, outer=\evenmargin\fi]{geometry}
\fi

\usepackage{bbding}
\usepackage{graphicx}
\makeatletter
\ifcolor
\RequirePackage[bookmarks,unicode,colorlinks=true]{hyperref}%
   \def\@citecolor{blue}%
   \def\@urlcolor{blue}%
   \def\@linkcolor{blue}%
\fi

\iforcid
\def\orcidID#1{\smash{\href{http://orcid.org/#1}{\protect\raisebox{-1.25pt}{\protect\includegraphics{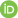}}}}}
\else
\def\orcidID#1{}
\fi
\makeatother

\ifcolor\fi

\usepackage[utf8]{inputenc}
\usepackage[T1]{fontenc}
\usepackage{xcolor}
\ifcolor\else\usepackage[hidelinks]{hyperref}\fi
\ifcolor\fi
\usepackage{subcaption}

\usepackage{algorithm}

\usepackage{mathtools}
\usepackage{amssymb}
\newcommand\doubleplus{+\kern-1.3ex+}
\DeclareUnicodeCharacter{2212}{\ensuremath{-}}  

\usepackage{listings}
\lstdefinestyle{alg}{%
  basicstyle=\sffamily\footnotesize,
  columns=fullflexible,
  morekeywords={let,match,if,else,in,then,with,function}
}
\lstdefinestyle{normal}{%
  basicstyle=\ttfamily\footnotesize,
  columns=flexible,
}
\lstdefinestyle{small}{%
  style=normal,
  basicstyle=\ttfamily\scriptsize,
}
\lstdefinestyle{numbers}{%
  numbers=left, showlines=true,
  frame=leftline,
  numberstyle=\tiny,
  numbersep=3pt,
  framexleftmargin=-2pt,
  xleftmargin=2em,
}
\lstdefinelanguage{RootPPL}[]{C++}{%
  morekeywords={SAMPLE,OBSERVE,INIT_MODEL,MAIN,ADD_BBLOCK,SMC,BBLOCK,PSTATE,NEXT,WEIGHT,BBLOCK_CALL,BBLOCK_JUMP}
}
\lstdefinelanguage{CorePPL}{%
  morekeywords={mexpr,let,assume,observe,true,false,include,type,con,in,lam,match,with,then,else,never,recursive,weight,resample,if},
  morecomment=[l]{--},
}
\newcommand{\rlstinline}{\lstinline[language=RootPPL]}
\newcommand{\clstinline}{\lstinline[language=CorePPL]}

\usepackage{tikz}
\usetikzlibrary{positioning,shapes,calc,patterns,automata,arrows.meta}

\usepackage{pgfplots}
\pgfplotsset{compat=1.15}
\usepgfplotslibrary{statistics}

\newcommand{\ttt}{\texttt}
\newcommand{\ra}{\rightarrow}
\newcommand{\term}{\textbf{\textsf{t}}}
\newcommand{\stmt}{\textbf{\textsf{stmt}}}
\newcommand{\tstmt}{\textbf{\textsf{tstmt}}}
\newcommand{\nextty}{\textbf{\textsf{next}}}
\newcommand{\accty}{\textbf{\textsf{acc}}}

\usepackage[absolute]{textpos}

\begin{document}

\title{%
  Compiling Universal Probabilistic Programming Languages with Efficient Parallel Sequential Monte Carlo Inference%
  \thanks{%
    This project is financially supported by the Swedish Foundation for Strategic Research (FFL15-0032 and RIT15-0012), the European Union's Horizon 2020 research and innovation program under the Marie Sk\l{}odowska-Curie grant agreement PhyPPL (No 898120), and the Swedish Research Council (grant number 2018-04620).
  }%
}
\titlerunning{Compiling Universal PPLs with Efficient Parallel SMC Inference}

\author{%
  Daniel Lundén\inst{1}(\raisebox{-2.4pt}{\Envelope})\orcidID{0000-0003-3127-5640} \and%
  Joey Öhman\inst{2}\orcidID{0000-0001-6342-268X} \and%
  Jan Kudlicka\inst{3}\orcidID{0000-0003-3806-4950} \and%
  Viktor Senderov\inst{4}\orcidID{0000-0003-3340-5963} \and%
  Fredrik Ronquist\inst{4,5}\orcidID{0000-0002-3929-251X} \and%
  David Broman\inst{1}\orcidID{0000-0001-8457-4105} \\[2mm]
}

\authorrunning{D. Lundén et al.}

\institute{%
  EECS and Digital Futures, KTH Royal Institute of Technology, Stockholm, Sweden, \email{\{dlunde,dbro\}@kth.se} \and
  AI Sweden, Stockholm, Sweden, \email{joey.ohman@ai.se} \and
  Department of Data Science and Analytics, BI Norwegian Business School, Oslo, Norway, \email{jan.kudlicka@bi.no} \and
  Department of Bioinformatics and Genetics, Swedish Museum of Natural History, Stockholm, Sweden, \email{\{viktor.senderov,fredrik.ronquist\}@nrm.se} \and
  Department of Zoology, Stockholm University, Stockholm, Sweden
}

\maketitle              

\ifauthor
\begin{textblock*}{1.2\textwidth}(1in + \oddsidemargin - 0.1\textwidth,10pt)
  \noindent
  \scriptsize
  This is an author-prepared version\ifextended, extended with appendices and minor additions to the main text\fi.
  © The Author(s) 2022.
  This version of the contribution\ifextended, except the extensions,\fi\ has been accepted for publication at ESOP 2022, after peer review, but is not the Version of Record (the version published by Springer) and does not reflect post-acceptance improvements, or any corrections.
  The Version of Record is available online at: \url{https://doi.org/10.1007/978-3-030-99336-8_2}.
\end{textblock*}
\fi

\begin{abstract}
  Probabilistic programming languages (PPLs) allow users to encode arbitrary inference problems, and PPL implementations provide general-purpose automatic inference for these problems.
  However, constructing inference implementations that are efficient enough is challenging for many real-world problems.
  Often, this is due to PPLs not fully exploiting available parallelization and optimization opportunities.
  For example, handling probabilistic \emph{checkpoints} in PPLs through continuation-passing style transformations or non-preemptive multitasking---as is done in many popular PPLs---often disallows compilation to low-level languages required for high-performance platforms such as GPUs.
  To solve the checkpoint problem, we introduce the concept of \emph{PPL control-flow graphs} (PCFGs)---a simple and efficient approach to checkpoints in low-level languages.
  We use this approach to implement \emph{RootPPL}: a low-level PPL built on CUDA and C++ with OpenMP, providing highly efficient and massively parallel SMC inference.
  We also introduce a general method of \emph{compiling} universal high-level PPLs to PCFGs and illustrate its application when compiling \emph{Miking CorePPL}---a high-level universal PPL---to RootPPL.
  The approach is the first to compile a universal PPL to GPUs with SMC inference.
  We evaluate RootPPL and the CorePPL compiler through a set of real-world experiments in the domains of phylogenetics and epidemiology, demonstrating up to 6$\times$ speedups over state-of-the-art PPLs implementing SMC inference.
\end{abstract}

\section{Introduction}\label{sec:intro}

\emph{Probabilistic programming languages} (PPLs) allow for encoding a wide range of statistical inference problems and provide \emph{inference algorithms} as part of their implementations.
Specifically, PPLs allow language users to focus solely on encoding their statistical problems, which the language implementation then solves automatically.
Many such languages exist and are applied in, e.g., statistics, machine learning, and artificial intelligence.
Some example PPLs are WebPPL~\cite{goodman2014design}, Birch~\cite{murray2018automated}, Anglican~\cite{wood2014new}, and Pyro~\cite{bingham2019pyro}.

However, implementing efficient PPL inference algorithms is challenging for many real-world problems.
Most often, \emph{universal}%
\footnote{%
   A term due to Goodman et al.~\cite{goodman2008church}. No precise definition exists, but in principle, a universal PPL program can perform probabilistic operations at any point. In particular, it is not always possible to statically determine the number of random variables.
} PPLs implement general-purpose inference algorithms---most commonly sequential Monte Carlo (SMC) methods \cite{doucet2001sequential}, Markov chain Monte Carlo (MCMC) methods \cite{gilks1995markov}, Hamiltonian Monte Carlo (HMC) methods \cite{carpenter2017stan}, variational inference (VI) \cite{wainwright2008graphical}, or a combination of these.
In some cases, poor efficiency may be due to an inference algorithm not well suited to the particular PPL program.
However, in other cases, the PPL implementations do not fully exploit opportunities for parallelization and optimization on the available hardware.
Unfortunately, doing this is often tricky without introducing complexity for end-users of PPLs.

A critical performance consideration is handling probabilistic \emph{checkpoints} \cite{tolpin2016design} in PPLs.
Checkpoints are locations in probabilistic programs where inference algorithms must interject, for example, to resample in SMC inference or record random draw locations where MCMC inference can explore alternative execution paths.
The most common approach to checkpoints---used in universal PPLs such as WebPPL \cite{goodman2014design}, Anglican \cite{wood2014new}, and Birch~\cite{murray2018automated}---is to associate them with PPL-specific language constructs.
In general, PPL users can place these constructs without restriction, and inference algorithms interject through continuation-passing style (CPS) transformations \cite{appel1991compiling,goodman2014design,wood2014new} or non-preemptive multitasking \cite{murray2018automated} (e.g., coroutines) that enable pausing and resuming executions.
These solutions are often not available in languages such as C and CUDA \cite{cuda2021} used for high-performance platforms such as graphics processing units (GPUs), making compiling PPLs to these languages and platforms challenging.
Some approaches for running PPLs on GPUs do exist, however.
LibBi~\cite{murray2013bayesian} runs on GPUs with SMC inference but is not universal.
Stan~\cite{carpenter2017stan} and AugurV2~\cite{huang2017compiling} partially run MCMC inference on GPUs but have limited expressive power.
Pyro~\cite{bingham2019pyro} runs on GPUs, but currently not in combination with SMC.
In this paper, we compile a universal PPL and run it with SMC on GPUs for the first time.

A more straightforward approach to checkpoints, used for SMC in Birch~\cite{murray2018automated} and Pyro~\cite{bingham2019pyro}, is to encode models with a \lstinline!step! function called iteratively.
Checkpoints then occur each time \lstinline!step! returns.
This paper presents a new approach to checkpoint handling, generalizing the \lstinline!step! function approach.
We write probabilistic programs as a set of code blocks connected in what we term a \emph{PPL control-flow graph} (PCFG).
PPL checkpoints are restricted to only occur at tail position in these blocks, and communication between blocks is only allowed through an explicit PCFG \emph{state}.
As a result, pausing and resuming executions is straightforward: it is simply a matter of stopping after executing a block and then resuming by running the next block.
A variable in the PCFG state, set from within the blocks, determines the next block.
This variable allows for loops and branching and gives the same expressive power as other universal PPLs.
We implement the above approach in \emph{RootPPL}: a low-level universal PPL framework built using C++ and CUDA with highly efficient and parallel SMC inference.
RootPPL consists of both an inference engine and a simple macro-based PPL.

\begin{figure}[tb]
\centering
\includegraphics[width=0.89\columnwidth]{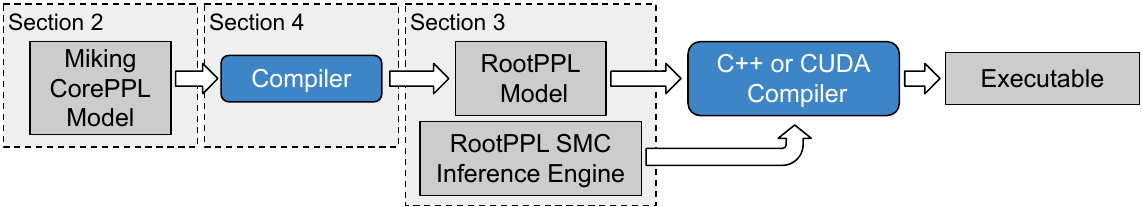}
\caption{%
  The CorePPL and RootPPL toolchain. Solid rectangular components (gray) represent programs and rounded components (blue) translations.
  The dashed rectangles indicate paper sections.
}
\label{fig:overview}
\end{figure}

A problem with RootPPL is that it is low-level and, therefore, challenging to write programs in.
In particular, sending data between blocks through the PCFG state can quickly get difficult for more complex models.
To solve this, we develop a general technique for \emph{compiling} high-level universal PPLs to PCFGs.
The key idea is to decompose functions in the high-level language to a set of PCFG blocks, such that checkpoints in the original function always occur at tail position in blocks.
As a result of the decomposition, the PCFG state must store a part of the call stack.
The compiler adds code for handling this call stack explicitly in the PCFG blocks.
We illustrate the compilation technique by introducing a high-level source language, \emph{Miking CorePPL}, and compiling it to RootPPL.
Fig.~\ref{fig:overview} illustrates the overall toolchain.

In summary, we make the following contributions.
\begin{itemize}
  \item
    We introduce PCFGs, a framework for checkpoint handling in PPLs, and use it to implement RootPPL: a low-level universal PPL with highly efficient and parallel SMC inference (Section~\ref{sec:pcfgrootppl}).

  \item
    We develop an approach for compiling high-level universal PPLs to PCFGs and use it to compile Miking CorePPL to RootPPL.
    In particular, we give an algorithm for decomposing high-level functions to PCFG blocks
    (Section~\ref{sec:compiling}).
\end{itemize}

Furthermore, we introduce Miking CorePPL in Section~\ref{sec:coreppl} and evaluate the performance of RootPPL and the CorePPL compiler in Section~\ref{sec:eval} on real-world models from phylogenetics and epidemiology, achieving up to 6$\times$ speedups over the state-of-the-art.
An artifact accompanying this paper supports the evaluation~\cite{lunden2022compilingartifact}.

\section{Miking CorePPL}\label{sec:coreppl}
This section introduces the Miking CorePPL language, used as a source language for the compiler in Section~\ref{sec:compiling}.
We discuss design considerations (Section~\ref{sec:cppldesign}) and present the syntax and semantics (Section~\ref{sec:cpplsyn}).

\subsection{Design Considerations}\label{sec:cppldesign}
Miking CorePPL (or CorePPL for short) is an \emph{intermediate representation} (IR) PPL, similar to IRs used by LLVM~\cite{llvm2021} and GCC~\cite{gcc2021}.
This allows the reuse of CorePPL as a target for domain-specific high-level PPLs and PPL compiler back-ends.
Consequently, CorePPL needs to be expressive enough to allow easy translation from various domain-specific PPLs and simple enough for practical use as a shared IR for compilers.
Therefore, we base CorePPL on the lambda calculus, extended with standard data types and constructs.

We must also consider which PPL-specific constructs to include.
Critically, most PPLs include constructs for defining random variables and likelihood updating \cite{gordon2014probabilistic}.
CorePPL includes such constructs, including first-class probability distributions, to match the expressive power of existing PPLs.

\subsection{Syntax and Semantics}\label{sec:cpplsyn}
We build CorePPL on top of the \emph{Miking} framework \cite{broman2019vision}: a meta-language system for creating domain-specific and general-purpose languages.
This allows reusing many existing Miking language components and transformations when building the CorePPL language.
More precisely, CorePPL extends \emph{Miking Core}---a core functional programming language in Miking---with PPL constructs.

A CorePPL program $\term$ is inductively defined by
\newcommand{\cl}{\clstinline}
\begin{equation}\label{eq:coreppl}
  \begin{aligned}
    \term \Coloneqq& \enspace
    x
    \enspace | \enspace
    \texttt{\cl!lam! $x$. \term}
    \enspace | \enspace
    \term_1 \enspace \term_2
    \enspace | \enspace
    \texttt{\cl!let! $x$ = $\term_1$ \cl!in! $\term_2$}
    \enspace | \enspace
    \texttt{$C$ $\term$}
    \enspace | \enspace
    \texttt{$c$}
    \\ |& \enspace
    \texttt{\cl!recursive! $[$\cl!let! $x$ = $\term$$]$ \cl!in!}
    \\ |& \enspace
    \texttt{\cl!match! $\term_1$ \cl!with! $p$ \cl!then! $\term_2$ \cl!else! $\term_3$}
    \enspace | \enspace
    \texttt{[$\term_1$, $\term_2$, $\ldots$, $\term_n$]}
    \\ |& \enspace
    \texttt{\{$l_1$ = $\term_1$, $l_2$ = $\term_2$, $\ldots$, $l_3$ = $\term_3$\}}
    \\ |& \enspace
    \texttt{\cl!assume! $\term$}
    \enspace | \enspace
    \texttt{\cl!weight! $\term$}
    \enspace | \enspace
    \texttt{\cl!observe! $\term_1$ $\term_2$}
    \enspace | \enspace
    \texttt{$D$ $\term_1$ $\term_2$ $\ldots$ $\term_{|D|}$}
  \end{aligned}
\end{equation}
where the metavariable $x$ ranges over a set of variable names; $C$ over a set of data constructor names; $p$ over a set of patterns; $l$ over a set of record labels; and $c$ over various literals, such as integers, floating-point numbers, booleans, and strings, as well as over various built-in functions in prefix form such as \lstinline!addi! (adds integers).
The notation $[$\cl!let! $x$ = $\term$$]$ indicates a sequence of mutually recursive \clstinline!let! bindings.
The metavariable $D$ ranges over a set of probability distribution names, with $|D|$ indicating the number of parameters for a distribution $D$.
For example, for the normal distribution, $|\mathcal{N}| = 2$.
In addition to \eqref{eq:coreppl}, we will also use the standard syntactic sugar \lstinline!;! to indicate sequencing, as well as \clstinline!if $\term_1$ then $\term_2$ else $\term_3$! for \clstinline!match $\term_1$ with true then $\term_2$ else $\term_3$!.

\begin{figure}[bt]
  \hspace{3mm}
  \begin{subfigure}{0.47\columnwidth}
    \centering
    \begin{tikzpicture}[node distance=4mm]
      \scriptsize

      \node[fill=black!30,minimum width=3cm,minimum height=4mm,anchor=north west]
        at (-2.25,0.15) {};

      \node (program) {%
          \begin{lstlisting}[style=normal,style=numbers,language=CorePPL]
recursive let geometric = lam p.
  let x = assume (Bernoulli p) in
  if x then
    weight (log 1.5);
    addi 1 (geometric p)
  else 1
in geometric 0.5
          \end{lstlisting}
        };
    \end{tikzpicture}
    \caption{}
    \label{fig:coreppl:a}
  \end{subfigure}
  \hfill
  \begin{subfigure}{0.35\columnwidth}
    \centering
    \begin{tikzpicture}[trim axis left, trim axis right]
      \pgfplotsset{title style={at={(0.6,0.35)}}}
      \begin{axis}[%
        title=Standard geometric,
        width=55mm,
        height=27mm,
        tick style={draw=none},
        ymin=0,
        ymax=0.55,
        xmax=11,
        xtick=\empty,
        bar width=9pt,
        ]
        \def\p{0.5}
        \def\w{1.0}
        \def\Z{(1/\w*((1/(1-\p*\w))-1))}
        \def\sumNine{(1/\w*(((1-(\p*\w)^10)/(1-\p*\w))-1))}
        \addplot [
          fill=gray,
          ybar,
          ] coordinates {%
            (1,  \p            / \Z)
            (2,  \p^2   * \w^1 / \Z)
            (3,  \p^3   * \w^2 / \Z)
            (4,  \p^4   * \w^3 / \Z)
            (5,  \p^5   * \w^4 / \Z)
            (6,  \p^6   * \w^5 / \Z)
            (7,  \p^7   * \w^6 / \Z)
            (8,  \p^8   * \w^7 / \Z)
            (9,  \p^9   * \w^8 / \Z)
            (10, \p^10  * \w^9 / \Z)
          };
        \Z
      \end{axis}
      \begin{axis}[%
        title=Weighted geometric,
        yshift=-12mm,
        width=55mm,
        height=27mm,
        tick style={draw=none},
        ymin=0,
        ymax=0.55,
        xmax=11,
        ylabel=Probability,
        xlabel=Outcome,
        xtick={1,2,0,3,4,5,6,7,8,9},
        y label style={at={(axis description cs:-0.15,1)},anchor=south},
        extra x ticks={10},
        extra x tick style={tick label style={xshift=1pt,yshift=-2.5pt}},
        extra x tick labels={$\cdots$},
        bar width=9pt,
        ]
        \def\p{0.5}
        \def\w{1.5}
        \def\Z{(1/\w*((1/(1-\p*\w))-1))}
        \def\sumNine{(1/\w*(((1-(\p*\w)^10)/(1-\p*\w))-1))}
        \addplot [
          fill=gray,
          ybar,
          ] coordinates {%
            (1,  \p            / \Z)
            (2,  \p^2   * \w^1 / \Z)
            (3,  \p^3   * \w^2 / \Z)
            (4,  \p^4   * \w^3 / \Z)
            (5,  \p^5   * \w^4 / \Z)
            (6,  \p^6   * \w^5 / \Z)
            (7,  \p^7   * \w^6 / \Z)
            (8,  \p^8   * \w^7 / \Z)
            (9,  \p^9   * \w^8 / \Z)
            (10, \p^10  * \w^9 / \Z)
          };
        \Z
      \end{axis}
    \end{tikzpicture}
    \caption{}
    \label{fig:coreppl:b}
  \end{subfigure}
  \hspace{3mm}
  \caption{%
    A toy example encoding a skewed geometric distribution, illustrating CorePPL. Part (a) gives the CorePPL program, and part (b) the corresponding distribution. The upper part of (b) shows the distribution for (a) with line 4 omitted, and the lower part of (b) shows it with line 4 included.
  }
  \label{fig:coreppl}
\end{figure}
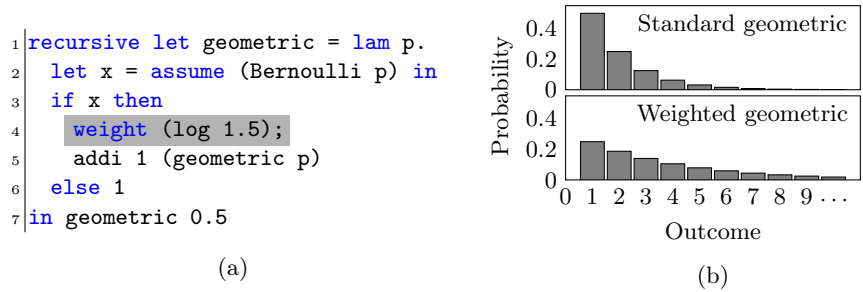
Consider the simple but illustrative CorePPL program in Fig.~\ref{fig:coreppl:a}.
The program encodes a variation of the geometric distribution, for which the result is the number of times a coin is flipped until the result is tails.
The program's core is the recursive function \lstinline!geometric!, defined using a function over the probability of heads for the coin, $p$.
We initially call this function at line 7 with the argument 0.5, indicating a fair coin.
On line 2, we define the random variable \lstinline!x! to have a Bernoulli distribution (i.e., a single coin flip) using the \clstinline!assume! construct (often known as \emph{sample} in PPLs with sampling-based inference).
If the random variable is \lstinline!false! (tails), we stop and return the result 1.
If the random variable is \lstinline!true! (heads), we keep flipping the coin by a recursive call to \lstinline!geometric! and add 1 to this result.
To illustrate likelihood updating, we make a contrived modification to the standard geometric distribution by adding \clstinline!weight (log 1.5)! on line 4.
This construct \emph{weights} the execution by a factor of 1.5 each time the result is heads.
Note that CorePPL weight computations are in log-space for numerical stability (hence the \clstinline!log 1.5! to factor by 1.5).
Thus, the unnormalized probability of seeing $n$ coin flips, including the final tails, is $0.5^n\cdot1.5^{n-1}$---where $1.5^{n-1}$ is the factor introduced by the $n-1$ calls to \clstinline!weight!.
The difference compared to the standard geometric distribution is illustrated in Fig.~\ref{fig:coreppl:b}.
The \clstinline!weight! construct is also commonly named \emph{factor} or \emph{score} in other PPLs.

What separates PPLs from ordinary programming languages is the ability to modify the likelihood of execution paths, akin to the use of \clstinline!weight! in Fig.~\ref{fig:coreppl:a}.
We often use likelihood modification to \emph{condition} a probabilistic model on observed data.
For this purpose, CorePPL includes an explicit \clstinline!observe! construct, which allows for modifying the likelihood based on observed data assumed to originate from a given probability distribution.
For instance, \clstinline!observe 0.3 (Normal 0 1)! updates the likelihood with $f_{\mathcal{N}(0,1)}(0.3)$ (note that this can equivalently be expressed through \clstinline!weight!), where $f_{\mathcal{N}(0,1)}$ is the probability density function of the standard normal distribution.
This conditioning can be related to Bayes' theorem: the random variables defined in a program define a prior distribution (e.g., the upper part of Fig.~\ref{fig:coreppl:b}), the use of the \clstinline!weight! and \clstinline!observe! primitives a likelihood function, and the inference algorithm of the PPL infers the posterior distribution (e.g., the lower part of Fig.~\ref{fig:coreppl:b})

CorePPL includes sequences, recursive variants, records, and pattern matching, standard in functional languages.
For example, \clstinline![1, 2, 3]! defines a sequence of length 3, \clstinline!{a = false, b = 1.2}! a record with labels \lstinline!a! and \lstinline!b!, and \clstinline!Leaf {age = 1.0}! a variant with the constructor name \clstinline!Leaf!, containing a record with the label \lstinline!age!.
The \clstinline!match! construct allows pattern matching.
For example, \clstinline!match a with Leaf {age = f} then f else 0.0!
checks if \lstinline!a! is a \lstinline!Leaf! and returns its age if so, or \clstinline!0.0! otherwise.
Here, \clstinline!f! is a pattern variable that is bound to the value of the \clstinline!age! element of \clstinline!a! in the \clstinline!then! branch of the \clstinline!match!.

The data types and pattern matching features in Miking, and consequently CorePPL, are not directly related to the paper's key contributions.
Therefore, we do not discuss them further.
However, the CorePPL compiler in Section~\ref{sec:cpplcompile} supports the features, and the CorePPL models in Section~\ref{sec:eval} make frequent use of them.
We consider CorePPL again in Section~\ref{sec:compiling} when compiling to PCFGs.

\section{PPL Control-Flow Graphs and RootPPL}\label{sec:pcfgrootppl}
This section introduces the new PCFG concept (Section~\ref{sec:pcfg}) and shows how to apply SMC over these (Section~\ref{sec:smcpcfg}). Finally, we present the PCFG and SMC-based RootPPL framework (Section~\ref{sec:rootppl}).

\subsection{PPL Control-Flow Graphs}\label{sec:pcfg}
In order to handle checkpoints efficiently without CPS or non-preemptive multitasking, we introduce \emph{PPL control-flow graphs} (PCFGs).
In contrast to traditional PPLs, where checkpoints are most often implicit, we make them explicit and central in the PCFG framework.
The main benefit of this approach is that the handling of checkpoints in inference algorithms is greatly simplified, which allows for implementing the framework in low-level languages.
However, the explicit checkpoint approach makes PCFGs relatively low-level, and they are mainly intended as a target when compiling from high-level PPLs.
We introduce such a compiler in Section~\ref{sec:compiling}.

Formally, we define a PCFG as a 6-tuple $(B,S,\mathit{sim},b_0,b_\text{stop},\mathcal{L})$.
The first component $B$ is a set of \emph{basic blocks} inspired by basic blocks used as a part of the control-flow analysis in traditional compilers \cite{aho2006compilers}.
In practice, the blocks in $B$ are pieces of code that together make up a complete probabilistic program.
Unlike basic blocks used in traditional compilers, we allow these pieces of code to contain branches internally.
The second component $S$ is a set of \emph{states}, representing collections of information that flow between basic blocks.
In practice, this state often contains local variables that live between blocks and an accumulated likelihood.
The blocks and states form the domain of the function $\mathit{sim}: B \times S \rightarrow B \times S \times \{\text{false},\text{true}\}$.
This function performs computation specific for the given block over the given state and outputs a \emph{successor} block indicating what to execute next, an updated state, and a boolean indicating whether or not there is a checkpoint at the end of the executed block.
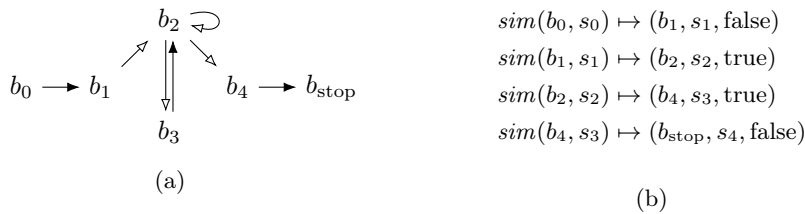
\begin{figure}[tb]
  \centering
  \begin{subfigure}[c]{0.55\textwidth}
    \centering
    \hspace{1cm}
    \normalsize
    \begin{tikzpicture}[node distance=5mm]
      \node (b0) {$b_0$};
      \node (b1) [right=of b0] {$b_1$};
      \node (b2) [above right=of b1] {$b_2$};
      \node (b3) [below right=of b1,yshift=3mm] {$b_3$};
      \node (b4) [below right=of b2] {$b_4$};
      \node (bend) [right=of b4] {$b_\text{stop}$};
      \node (end) [right=of bend] {};

      \draw[-{Latex[]}] (b0) to (b1);
      \draw[-{Latex[open]}] (b1) to (b2);
      \draw[-{Latex[open]}] (b2) to[loop right] (b2);
      \draw[-{Latex[open]}] (b2.260) to (b3.100);
      \draw[-{Latex[open]}] (b2) to (b4);
      \draw[-{Latex[]}] (b3.80) to (b2.280);
      \draw[-{Latex[]}] (b4) to (bend);
    \end{tikzpicture}
    \caption{}
    \label{fig:pcfg}
  \end{subfigure}
  \hfill
  \begin{subfigure}[c]{0.4\textwidth}
    \[
      \begin{aligned}
        \mathit{sim}(b_0,s_0)& \mapsto (b_1,s_1,\text{false})           \\
        \mathit{sim}(b_1,s_1)& \mapsto (b_2,s_2,\text{true})            \\
        \mathit{sim}(b_2,s_2)& \mapsto (b_4,s_3,\text{true})            \\
        \mathit{sim}(b_4,s_3)& \mapsto (b_\text{stop},s_4,\text{false}) \\
      \end{aligned}
    \]
    \caption{}
    \label{fig:pcfgexample}
  \end{subfigure}
  \caption{%
    A PCFG illustration.
    Part (a) shows an example PCFG.
    The arrows denote the possible flows of control between the blocks, with regular arrows denoting checkpoint transitions and arrows with open tips non-checkpoint transitions.
    Part (b) shows a possible execution sequence with $\mathit{sim}$ for (a).
  }
\end{figure}

To illustrate this formalization, consider the PCFG in Fig.~\ref{fig:pcfg} for which $B = \{b_0,b_1,\ldots,b_4,b_\text{stop}\}$.
The block $b_0$ is present in every PCFG and represents its entry point.
Similarly, the block $b_\text{stop}$ is a unique block indicating termination, which must be reachable from all other blocks.
For some initial state $s_0 \in S$, Fig.~\ref{fig:pcfgexample} illustrates a possible execution sequence starting at $b_0$ in Fig.~\ref{fig:pcfg} before terminating at $b_\text{stop}$.
The structure of a PCFG restricts checkpoints to \emph{only} occur at the end of basic blocks and confines communication between blocks to the state.
These restrictions greatly simplify inference algorithm implementations.
More precisely, rather than relying on CPS or non-preemptive multitasking, the inference algorithm can simply run a block $b$ with $\mathit{sim}$, handle the checkpoint, and then run the successor block indicated by the output of $\mathit{sim}$.


\subsection{SMC and PCFGs}\label{sec:smcpcfg}
\begin{algorithm}[tb]
  \caption{%
    A standard SMC algorithm applied to PCFGs.
  }%
  \label{alg:smc}
  \textbf{Input:} A PCFG $(B,S,\mathit{sim},b_0,b_\text{stop},\mathcal{L})$. A set of initial states $\{s_n\}_{n=1}^N$. \\
  \textbf{Output:} An updated set of states $\{s_n\}_{n=1}^N$.
  \begin{enumerate}

    \item \textbf{Initialization:}
      For each $1 \leq n \leq N$, let $a_n \coloneqq b_0$ and $c_n \coloneqq \text{false}$.

    \item \textbf{Propagation:}
      \label{alg:smc:prop}
      If all $a_n = b_\text{stop}$, terminate and output $\{s_n\}_{n=1}^N$.
      If not, for each $1 \leq n \leq N$ where $c_n = \text{false}$, let $(a_n,s_n,c_n) \coloneqq \mathit{sim}(a_n,s_n)$.
      If all $c_n = \text{true}$, go to~\ref{alg:smc:resampling}.
      If not, repeat~\ref{alg:smc:prop}.

    \item \textbf{Resampling:}
      \label{alg:smc:resampling}
      For each $1 \leq n \leq N$, let $p_n \coloneqq \mathcal{L}(s_n) / \sum_{i=1}^N \mathcal{L}(s_i)$.
      For each $1 \leq n \leq N$, draw a new index $i$ from $\{i\}_{i=1}^N$ with probabilities $\{p_i\}_{i=1}^N$. Let $(s'_n,b'_n) \coloneqq (s_i,b_i)$. Finally, for each $1 \leq n \leq N$, let $(s_n,b_n,c_n) \coloneqq (s'_n,b'_n,\text{false})$.
      Go to~\ref{alg:smc:prop}.

  \end{enumerate}
\end{algorithm}

To prepare for introducing RootPPL in Section~\ref{sec:rootppl}, we present how to apply SMC inference to PCFGs.
The work by Naesseth et al.~\cite{naesseth2019elements} contains a more general and pedagogical introduction to SMC.
At a high level, SMC inference works by simulating many instances---known as \emph{particles} in SMC literature---of a PCFG program concurrently, occasionally \emph{resampling} the different particles based on their current likelihoods.
In CorePPL, for example, such likelihoods are determined by \clstinline!weight! and \clstinline!observe!.
Resampling allows the downstream simulation to focus on particles with a higher likelihood.

In order to apply SMC inference over PCFGs, we need some way of determining the likelihood of the SMC particles.
For this, we use the final component of the PCFG definition, $\mathcal{L}: S \rightarrow \mathbb{R}_{\geq0}$, which is a function mapping states to a likelihood (a non-negative real number).
Concretely, this likelihood is most often stored directly in the state as a real number, and $\mathcal{L}$ simply extracts it.

Algorithm~\ref{alg:smc} defines an SMC algorithm over PCFGs.
It takes a PCFG as input, together with a set of $N$ states $\{s_n\}_{n=1}^N$, which represent the SMC particles.
Step 1 in the algorithm sets up variables $a_n$ and $c_n$, indicating for each particle its current block and whether or not a checkpoint has occurred in it.
Step 2 simulates all particles that have not yet reached a checkpoint using $\mathit{sim}$.
This step repeats until all particles have reached a checkpoint (this is a synchronization point for parallel implementations).
Step 3 uses the likelihood function $\mathcal{L}$ to compute the relative likelihoods of all particles and then \emph{resamples} them based on this.
That is, we sample $N$ particles from the existing $N$ particles (with replacement) based on the relative likelihoods.
After resampling, we return to step 2.
If all particles have reached the termination block $b_\text{stop}$, the algorithm terminates and returns the current states.

Note in Algorithm~\ref{alg:smc} that the input states are \emph{not} required to be identical.
For example, each state should have a unique seed used to generate random numbers (e.g., with \clstinline!assume! in CorePPL).
Non-identical initial states in Algorithm~\ref{alg:smc} imply that different particles may traverse the blocks in $B$ differently and reach checkpoints at different times.
Although this means that different particles can be at different blocks concurrently, the SMC algorithm is still correct \cite{lunden2021correctness}.
This PCFG property is essential as it allows for the encoding of universal probabilistic programs in PCFG-based PPLs.
Furthermore, it implies that some particles may reach $b_\text{stop}$ earlier than others.
To solve this, we require in Algorithm~\ref{alg:smc} that $\mathit{sim}(b_\text{stop},s) = (b_\text{stop},s,\text{true})$ holds for all states $s$.
That is, particles that have finished also participate in resampling and cannot cause step 2 to loop infinitely.

Next, we describe our implementation of PCFGs with SMC: RootPPL.

\subsection{RootPPL}\label{sec:rootppl}
We make use of the PCFG framework when implementing RootPPL:
a new low-level PPL framework built on top of CUDA C++ and C++, intended for highly optimized and massively parallel SMC inference on general-purpose GPUs.
RootPPL consists of two major components: a macro-based C++ PPL for encoding probabilistic models and an SMC inference engine.

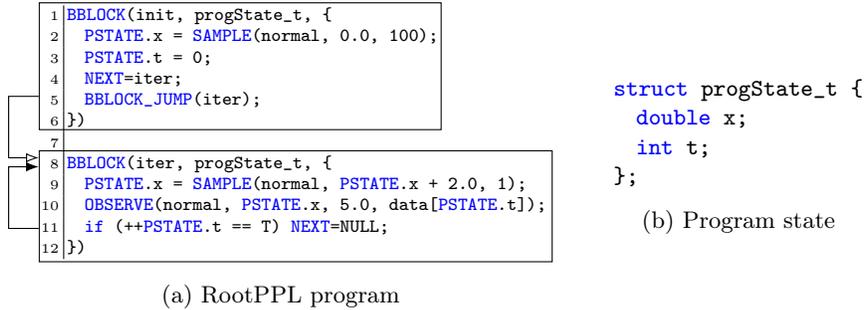
\begin{figure}[tb]
  \begin{subfigure}[c]{0.65\columnwidth}
    \centering
    \begin{tikzpicture}[node distance=4mm]
      \scriptsize
      \node (program) {%
          \hspace{6pt}
          \begin{lstlisting}[style=small,style=numbers,language=RootPPL]
BBLOCK(init, progState_t, {
  PSTATE.x = SAMPLE(normal, 0.0, 100);
  PSTATE.t = 0;
  NEXT=iter;
  BBLOCK_JUMP(iter);
})

BBLOCK(iter, progState_t, {
  PSTATE.x = SAMPLE(normal, PSTATE.x + 2.0, 1);
  OBSERVE(normal, PSTATE.x, 5.0, data[PSTATE.t]);
  if (++PSTATE.t == T) NEXT=NULL;
})
          \end{lstlisting}
        };

      \node[draw,anchor=north west,
        minimum width=152pt,minimum height=48pt,
        xshift=-97pt,yshift=48pt] (b1) {};

      \node[draw,anchor=north west,
        minimum width=194pt,minimum height=42pt,
        xshift=-97pt,yshift=-8pt] (b2) {};

      \draw[-{Latex[open]}]
        ($(b1.south west)+(0,0.45)$)
        -- ++(-0.4,0)
        |- ($(b2.north west)-(0,0.10)$);

      \draw[-{Latex[]}]
        ($(b2.south west)+(0,0.45)$)
        -- ++(-0.4,0)
        |- ($(b2.north west)-(0,0.22)$);

    \end{tikzpicture}
    \caption{RootPPL program}
    \label{fig:rootppl:a}
  \end{subfigure}
  \begin{subfigure}[c]{0.33\columnwidth}
    \centering
    \begin{tabular}{c}
      \begin{lstlisting}[style=normal,language=RootPPL]
struct progState_t {
  double x;
  int t;
};
      \end{lstlisting}
    \end{tabular}
    \caption{Program state}
    \label{fig:rootppl:b}
  \end{subfigure}

  \caption{%
    Part (a) illustrates a RootPPL program encoding the state-space model in \eqref{eq:ssm}.
    The text provides details.
    We set \rlstinline!NEXT! at line 4 rather than in \rlstinline!iter! as an optimization.
    Part (b) defines the RootPPL program state type \lstinline!progState_t!.
  }
  \label{fig:rootppl}
\end{figure}

The macro-based language has two purposes: to support compiling the same program to either CPU or GPU and to simplify the encoding of models for programmers.
As a result, the macros hide all hardware details from the programmer.
To illustrate this macro-based PPL, consider the example RootPPL program in Fig.~\ref{fig:rootppl:a}.
This program encodes a simple state-space model for an object moving along an axis in $\mathbb{R}$, given by
\begin{equation}\label{eq:ssm}
  X_0 \sim \mathcal{N}(0,100), \quad
  X_t \sim \mathcal{N}(X_{t-1} + 2, 1), \quad
  Y_t \sim \mathcal{N}(X_t, 5), \quad
  1 \leq t \leq T.
\end{equation}
Here, $X_0$ is the initial position, $X_t$ the following positions, and $Y_t$ a set of noisy observations of the object position.
The inference goal is to determine the distribution of $X_T$ (the final position of the object) conditioned on all $Y_t$.

Fig.~\ref{fig:rootppl:a} implements~\eqref{eq:ssm} with two basic blocks, introduced with the \rlstinline!BBLOCK! macro in RootPPL.
The first block \lstinline!init! draws $X_0$ using the \rlstinline!SAMPLE! macro (equivalent to \clstinline!assume! in CorePPL) on line 2 and stores the drawn value in the \emph{program state} variable \lstinline!x! through the \rlstinline!PSTATE! macro.
This program state is the RootPPL instantiation of the PCFG state introduced in Section~\ref{sec:pcfg}.
Another program state variable, \lstinline!t! (corresponding to the index $t$ in the model), is initialized on line 3.
As preparation for iterating over the \lstinline!iter! block, we set the \rlstinline!NEXT! construct to \lstinline!iter! at line 4.
Finally, the block exits by making a direct non-checkpoint transition to \rlstinline!iter! using the \rlstinline!BBLOCK_JUMP! macro at line 5.

In \lstinline!iter!, we sample $X_1$ at line 9 and write the result to \lstinline!x! (overwriting the previous $X_0$, which is no longer needed).
Line 10 updates the likelihood using the \rlstinline!OBSERVE! macro (equivalent to \clstinline!observe! in CorePPL), corresponding to observing $Y_1$ in the model.
We access all $Y_t$ through the \lstinline!data! array, a shared global constant, avoiding memory duplication in the program state.
Finally, at line 11, we check if we are at time \lstinline!T! (a shared global constant for $T$).
If this is the case, \rlstinline!NEXT! is set to \lstinline!NULL!, indicating termination.
This is equivalent to moving to $b_\text{stop}$ in the PCFG formalization.
Otherwise, \rlstinline!NEXT! keeps its value set at line 4 and jumps to the beginning of the \lstinline!iter! block.
Not using \rlstinline!BBLOCK_JUMP! allows \lstinline!iter! to return to the inference engine between iterations, indicating checkpoint transitions.
In RootPPL, this means that SMC inference will resample the instances before returning to \lstinline!iter! for the next iteration.

The programmer defines the RootPPL program state for each RootPPL program as an arbitrary C++ struct type and passes
this type (e.g., \rlstinline!progState_t! in Fig.~\ref{fig:rootppl:a}) to each basic block.
The \rlstinline!PSTATE! macro accesses the variables in the struct.
Fig.~\ref{fig:rootppl:b} illustrates the program state for the example program in Fig.~\ref{fig:rootppl:a}.
As described in Section~\ref{sec:pcfg}, this program state is the \emph{only} possible means to pass data from one basic block to another in RootPPL.

This minimal example does not illustrate all RootPPL language features (e.g., \rlstinline!WEIGHT!).
Further details on the RootPPL language are available at GitHub \cite{mikingdppl2021}.

The second part of the RootPPL framework is the SMC inference engine.
It is crucial to take advantage of the highly parallel nature of SMC and available hardware for parallelization to achieve high performance.
For this purpose, RootPPL supports compilation to either C++ on single-core, C++ on multicore through OpenMP \cite{openmp2021}, and CUDA C++ \cite{cuda2021} with massive parallelism on the GPU.

We present the main inference loop in RootPPL below (cf. Algorithm~\ref{alg:smc}).
\begin{enumerate}
  \item Initialize random seeds.
  \item Execute the basic block indicated by \rlstinline!NEXT! for all particles. This execution may include a chain of blocks with non-checkpoint transitions between them (using the \rlstinline!BBLOCK_JUMP! macro) before returning to the inference engine.
  \item If all particles have terminated (i.e., \rlstinline!NEXT = NULL!), stop.
  \item Resample all particles and go to 2.
\end{enumerate}
The random seeds in step 1 are initialized differently depending on the compile target.
For plain C++ on a single core, one seed is shared between all particles because they are executed sequentially.
However, for OpenMP and CUDA, the parallel execution requires that we assign each thread a unique seed shared between all particles running on it.
For CUDA, these seeds are placed in thread-local CUDA memory for each particle to minimize memory overhead when using \rlstinline!SAMPLE! (which is performance-critical).
In addition, when compiling to CUDA, we initialize the seeds in parallel using a CUDA compute kernel.

Step 2 executes the particles sequentially, in parallel using OpenMP threads, or in parallel using a CUDA compute kernel.
Step 3 then performs a termination check.
First, we check if the first particle has terminated.
If it has not terminated, we directly move to the resampling step.
If it has terminated, we iteratively check other particles to either find a particle that has not terminated or conclude that all particles have terminated and stop the inference.
This approach both allows for particles terminating at different times and introduces minimal overhead for the case when all particles terminate simultaneously (which is quite common).
When all particles terminate simultaneously, it is enough to check the first particle in all iterations of step 3 except the last.

The resampling step is the most difficult one to parallelize efficiently.
The reason is the normalizing sum (e.g., $\sum_{i=1}^N \mathcal{L}(s_i)$ in Algorithm~\ref{alg:smc}) that we must compute in order to determine resampling probabilities.
We use systematic resampling for single-core and OpenMP and parallel systematic resampling for CUDA, as described in Murray et al.~\cite{murray2016parallel} (we do not use in-place propagation).
We compute the normalizing sum in parallel via the Thrust library \cite{thrust2021} for CUDA.

Another important consideration for the inference engine is memory allocation.
In particular, the memory allocated for \rlstinline!NEXT!, the likelihood, and the \rlstinline!PSTATE! for each particle, is laid out as separate arrays in memory, rather than one big array of structs.
This approach, known as memory coalescing, avoids strided memory accesses in global memory and is preferred for parallel operations, particularly for CUDA.
Another memory consideration is particle duplication during resampling.
For this, we use a custom aligned memory transfer in CUDA because the standard \lstinline!memcpy! implementation in CUDA proved to be a bottleneck.
With a single core and OpenMP, \lstinline!memcpy! runs without issue.
Additionally, we perform a specific optimization when copying the program state used in the CorePPL compiler.
This program state consists of a possibly large stack (with user-definable size) together with a stack pointer, and we ensure not to copy the unused part of the stack located beyond the stack pointer.
This is a critical optimization for the CorePPL compiler.

Other things supported in RootPPL are the estimation of \emph{normalizing constants} for encoded models and adaptive resampling based on the current \emph{effective sample size} (ESS).
These are standard concepts in SMC inference.
For more details, see, e.g., Naesseth et al.~\cite{naesseth2019elements}.

Next, we use RootPPL as the target language for the CorePPL compiler.

\section{Compiling to PCFGs}\label{sec:compiling}
This section introduces the ideas for compiling high-level universal PPLs to PCFGs.
We present the key transformation---\emph{function decomposition} into basic blocks---using a toy example (Section~\ref{sec:decex}), a formal algorithm (Section~\ref{sec:decalg}),
a high-level overview of the CorePPL-to-RootPPL compiler (Section~\ref{sec:cpplcompile}), and the compilers strengths and limitations (Section~\ref{sec:lim}).

\begin{figure}[tbp]
  \centering
  \begin{subfigure}[c]{0.4\columnwidth}
    \centering
\begin{lstlisting}[style=small,style=numbers,firstline=3,lastline=18,language=CorePPL]
recursive let f: Float -> Float =
 lam p.
  let s1 = assume (Gamma p p) in
  resample;
  let s2 =
    if geqf s1 1. then 2.
    else 3. in
  let s3 =
    if leqf s2 4. then
      let s4 =
        if eqf s2 5. then 6.
        else f 7. in
      addf s4 s4
    else 8. in
  mulf s3 s3
in
\end{lstlisting}
    \ifthesis\vspace{-3mm}\fi
    \caption{Source CorePPL program.}
    \label{fig:example:a}
  \end{subfigure}%
  \begin{subfigure}[c]{0.45\columnwidth}
    \centering
    \begin{tikzpicture}[node distance=4mm]
      \scriptsize
      \node (program) {%
          \hspace{6pt}
          \begin{lstlisting}[style=small,style=numbers,language=CorePPL]
recursive let f: Float -> Float =
 lam p.
  let s1 = assume (Gamma p p) in
  resample;
  let t1 = geqf s1 1.  in
  let s2 = if t1 then 2. else 3. in
  let t2 = leqf s2 4. in
  let s3 =
    if t2 then
      let t3 = eqf s2 5.  in
      let s4 =
        if t3 then 6. else f 7. in
      addf s4 s4
    else 8. in
  mulf s3 s3
in
          \end{lstlisting}

        };

      \node[draw,anchor=north west,
        minimum width=115pt,minimum height=15.5pt,
        xshift=-55pt,yshift=47pt] (b1) {};
      \node[anchor=west,inner sep=1pt] at (b1.east) {1};

      \node[draw,anchor=north west,
        minimum width=125pt,minimum height=79.5pt,
        xshift=-55pt,yshift=31pt] (b2) {};
      \node[anchor=west,inner sep=1pt] at (b2.east) {2};

      \node[draw,anchor=north west,
        minimum width=39pt,minimum height=8pt,
        xshift=-41pt,yshift=-33pt] (b3) {};
      \node[anchor=west,inner sep=1pt] at (b3.east) {3};

      \node[draw,anchor=north west,
        minimum width=40pt,minimum height=8pt,
        xshift=-55pt,yshift=-49pt] (b4) {};
      \node[anchor=west,inner sep=1pt] at (b4.east) {4};

    \end{tikzpicture}
    \ifthesis\vspace{-2mm}\fi
    \caption{Intermediate ANF representation.}
    \label{fig:example:b}
  \end{subfigure}

  \begin{subfigure}[c]{\columnwidth}
    \scriptsize
    \centering
    {
      \lstset{basicstyle=\ttfamily\scriptsize}
      \begin{tikzpicture}[node distance=4mm]

        \node[draw]
          (b1) {%
            \hspace{4pt}
            \begin{lstlisting}[style=small,style=numbers,language=RootPPL]
struct STACK_f *sf =
  PSTATE.stack
  + PSTATE.stackPtr
  - sizeof(struct STACK_f);
sf->s1 =
  SAMPLE(gamma, sf->p, sf->p);
NEXT = 2;
            \end{lstlisting}
          };
        \node[anchor=south west,inner xsep=0] at (b1.north west) {1};

        \node[%
          draw,right=of b1.north east,anchor=north west,yshift=16pt
          ] (b2) {%
            \hspace{7pt}
            \begin{lstlisting}[style=small,style=numbers,language=RootPPL]
struct STACK_f *sf = $\ldots$;
char t1 = sf->s1 >= 1.;
double s2;
if (t1 == 1) { s2 = 2.; }
else { s2 = 3.; }
char t2 = s2 <= 4.;
if (t2 == 1) {
  char t3 = s2 == 5.;
  if (t3 == 1) {
    sf->s4 = 6.;
    BBLOCK_JUMP(3);
  } else {
    struct STACK_f *callsf =
      PSTATE.stack
      + PSTATE.stackPtr;
    callsf->ra = 3;
    callsf->p = 7.;
    callsf->retValLoc =
      &(sf->s4)
      - PSTATE.stack;
    PSTATE.stackPtr =
      PSTATE.stackPtr
      + sizeof(struct STACK_f);
    BBLOCK_JUMP(1);
  }
} else {
  sf->s3 = 8.;
  BBLOCK_JUMP(4);
}
            \end{lstlisting}
          };
        \node[anchor=north west,inner xsep=0,xshift=0.3mm] at (b2.north east) {2};

        \node[draw,below=20pt of b1.south east,anchor=north east] (b3) {%
            \hspace{4pt}
            \begin{lstlisting}[style=small,style=numbers,language=RootPPL]
struct STACK_f *sf = $\ldots$;
sf->s3 = sf->s4 + sf->s4;
BBLOCK_JUMP(4);
            \end{lstlisting}
          };
        \node[anchor=south west,inner xsep=0] at (b3.north west) {3};

        \node[draw,below=20pt of b3.south east,anchor=north east] (b4) {%
            \hspace{4pt}
            \begin{lstlisting}[style=small,style=numbers,language=RootPPL]
struct STACK_f *sf = $\ldots$;
double t = sf->s3 * sf->s3;
*(PSTATE.stack + sf->retValLoc) = t;
PSTATE.stackPtr =
  PSTATE.stackPtr
  - sizeof(struct STACK_f);
BBLOCK_JUMP(sf->ra);
            \end{lstlisting}
          };
        \node[anchor=south west,inner xsep=0] at (b4.north west) {4};

        \node (ra) at ($(b4.south)+(0,-0.5)$) {};
        \node (init) at ($(b1.north)+(0,0.5)$) {};

        \draw[-{Latex[]}]
          ($(b1.south east)+(0,0.2)$)
          -- ++(0.15,0)
          |- ($(b2.north west)-(0,0.3)$);
        \draw[-{Latex[open]}]
          ($(b2.south west)+(0,1.65)$)
          -- ++(-0.1,0)
          -- ++(0,3.75)
          -- ++(-4.85,0)
          |- ($(b1.north west)+(0,-0.35)$);
        \draw[-{Latex[open]}]
          ($(b2.south west)+(0,5.29)$)
          -- ++(-0.06,0)
          arc[start angle=0, end angle=180,radius=0.04]
          -- ++(-0.06,0)
          |- ($(b3.north east)+(0,-0.33)$);
        \draw[-{Latex[open]}]
          ($(b3.south east)+(0,0.2)$)
          -- ++(0.2,0)
          |- ($(b4.north east)+(0,-0.33)$);
        \draw[-{Latex[open]}]
          ($(b2.south west)+(0,0.55)$)
          -- ++(-0.2,0)
          |- ($(b4.north east)+(0,-0.33)$);
        \draw[-{Latex[open]}] (b4) to (ra);
        \draw[-{Latex[open]}] (init) to (b1);
      \end{tikzpicture}
    }
    \ifthesis\vspace{-1mm}\fi
      \caption{%
        Compiled RootPPL PCFG illustration.
        Some RootPPL constructs are omitted or slightly modified for readability.
        In particular, we omit the \rlstinline{BBLOCK} construct used in Fig.~\ref{fig:rootppl:a}.
        Instead, we illustrate the blocks as nodes in a graph, numbered by indices.
        The arrows indicate control flow between the blocks, with the incoming arrow to block 1 representing the call to \lstinline!f! and the outgoing arrow from block 4 representing the return from \lstinline!f!.
      }
    \label{fig:example:c}
  \end{subfigure}
  \ifthesis\vspace{-2mm}\fi
  \caption{%
    Compilation of a CorePPL program (a) to a RootPPL PCFG (c).
    Part (b) illustrates an intermediate ANF representation of (a) and also indicates the parts of the program corresponding to the blocks in (c).
    We provide further details in the text.
  }
  \label{fig:example}
\end{figure}
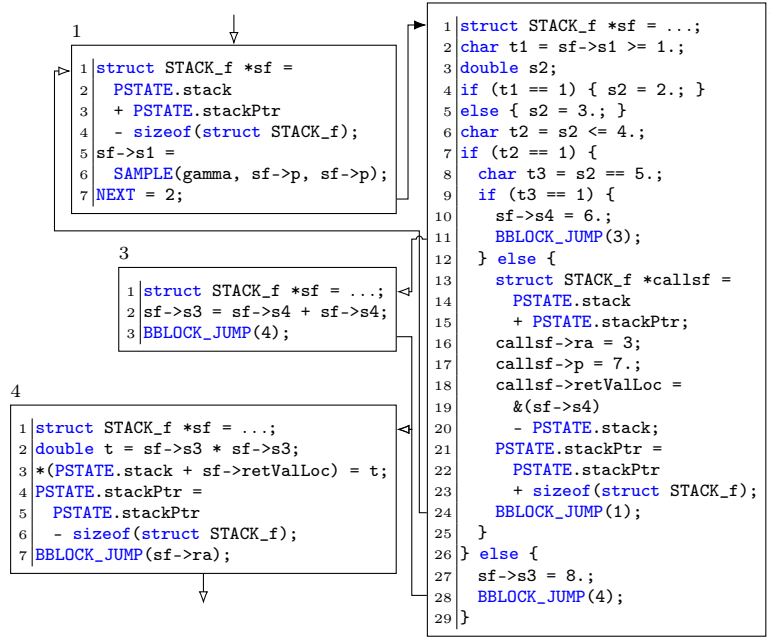

\subsection{Function Decomposition Example}\label{sec:decex}
The major challenge when compiling high-level PPLs is implementing pausing and resuming at checkpoints to yield control to an inference algorithm temporarily.
Pausing and resuming in low-level languages is especially difficult due to runtime limitations.
We solve this problem by compiling to the PCFGs introduced in Section~\ref{sec:pcfgrootppl}, specifically designed for implementation in low-level target languages.
A challenge with this approach is that checkpoints can occur at arbitrary locations in high-level probabilistic programs, whereas in PCFGs, checkpoints must always occur at tail position in basic blocks.
We solve this by \emph{decomposing} functions in the source language into a set of basic blocks.
Our approach is similar to how functions are decomposed into basic blocks in standard compilers such as GCC \cite{gcc2021} and LLVM \cite{llvm2021} (see, e.g., Aho et al.~\cite{aho2006compilers}).
The difference is that we only decompose \emph{as needed}, based on where checkpoints occur.
In particular, we do \emph{not} decompose functions, and parts of functions, in which checkpoints are guaranteed not to occur.
This allows for more optimizations by the underlying compiler (e.g., NVCC or GCC for RootPPL).

Consider the toy CorePPL function in Fig.~\ref{fig:example:a} and the resulting compilation to a RootPPL PCFG in Fig.~\ref{fig:example:c}.
For this example, we introduce an explicit SMC checkpoint \clstinline!resample! in CorePPL, indicating where SMC should pause executions in order to resample.
The \clstinline!resample! construct is the sole checkpoint considered in this example (and the CorePPL compiler), but the method generally applies for arbitrary checkpoints.
Optimally, the \clstinline!resample! construct should be automatically inserted by the compiler \cite{lunden2018automatic}.
However, we do not consider this problem in this paper and assume \clstinline!resample!s are inserted prior to compilation.
The first step in the decomposition is to translate the program into A-normal form (ANF) \cite{flanagan1993essence}, illustrated in Fig.~\ref{fig:example:b}.
ANF is commonly used in compilers and ensures that non-trivial expressions (e.g., function applications and checkpoints) are always name-bound.
For CorePPL, ANF guarantees that the body of each \clstinline!let! expression, or expression in tail position, is trivial, contains at most one function application, or is an \clstinline!if! expression with a trivial condition, resulting in simplified decomposition. We will use the program in Fig.~\ref{fig:example:b} as the target for decomposition in the following.
Note that variables introduced by ANF start with a \lstinline!t! in Fig.~\ref{fig:example:b}, while the original variables from Fig.~\ref{fig:example:a} start with an \lstinline!s!.

The goal with the decomposition is to ensure that we \emph{immediately} return control to the inference engine at checkpoints.
In the PCFG framework, the only way to fulfill this is to ensure that checkpoints occur at tail position in basic blocks.
First, consider the \clstinline!resample! checkpoint at line 4 in Fig.~\ref{fig:example:b}, causing a split into blocks 1 and 2 in the compiled RootPPL PCFG in Fig.~\ref{fig:example:c}.
Note that in block 1, \rlstinline!NEXT! is set to 2 at line 7 before returning, indicating that the inference engine should resume execution at block 2 after handling the checkpoint, also illustrated by a closed arrow.
Note the stack frame pointer \rlstinline!sf! in block 1 for this invocation of \lstinline!f!, which points to a location in an explicit call stack in the RootPPL program state \rlstinline!PSTATE!.
We require such a call stack due to compiling to PCFGs---\emph{any} data that lives between basic blocks (e.g., a call stack), such as \lstinline!s1!, \emph{must} be put in the program state.
We define the stack frame pointer \rlstinline!sf! equivalently at the top of all blocks for the decomposed function \clstinline!f! in Fig.~\ref{fig:example:c} but replace the definition with $\ldots$ in blocks other than the first for brevity.

It is not sufficient to split into blocks at explicit checkpoints.
Consider, for example, the recursive call to \lstinline!f! in the \clstinline!else! branch on line 12 in Fig.~\ref{fig:example:b}.
During this function call, we encounter at least one \clstinline!resample!, resulting in at least one block split within the function,
meaning that all data required by \lstinline!f! must be put in an explicit stack frame and stored in the program state. If not, we lose the data between the basic blocks of \lstinline!f!.
In particular, the block return address \lstinline!ra! is stored in the stack frame, indicating which block to return to at the end of the function call.
In the case of the call to \lstinline!f! at line 12 in Fig.~\ref{fig:example:b}, we must return to line 13.
Therefore, we must place line 13 at the beginning of a basic block in Fig.~\ref{fig:example:c} (block 3).
In general, we must place all calls to decomposed functions (i.e., functions that may, directly or indirectly, encounter a checkpoint) at tail position in basic blocks.
Besides line 13 in Fig.~\ref{fig:example:b}, this also means that line 15 in Fig.~\ref{fig:example:b} cannot be part of block 2.
It cannot be part of block 3 either because it may be executed independently of line 13 in Fig.~\ref{fig:example:b} if we take the \clstinline!else! branch of the \clstinline!if! at line 9 in Fig.~\ref{fig:example:b}.
Consequently, we must put it in a separate block (block 4 in Fig.~\ref{fig:example:c}).
The decomposition of function applications and \clstinline!if! expressions is similar to how standard compilers decompose machine instructions into basic blocks (sequences of instructions without any internal jumps or branches) \cite{aho2006compilers}.
The difference, however, is that we do not split into blocks at \emph{all} \clstinline!if! expressions and function calls.
For example, the \clstinline!if! at line 6 in Fig.~\ref{fig:example:b} is guaranteed not to include a checkpoint and can be left untouched (lines 4--5 in Fig.~\ref{fig:example:c}).
Similarly, the call to \lstinline!geqf! at line 5 in Fig~\ref{fig:example:b} is guaranteed not to encounter any checkpoints.
Conservatively determining which functions are guaranteed not to encounter any checkpoints can be done through static analysis.
Such a static analysis phase is part of the CorePPL compiler, described in Section~\ref{sec:cpplcompile}.

We now take a closer look at the call stack handling in Fig.~\ref{fig:example:c}.
The following description is specific for RootPPL, but similar solutions must be applied if compiling to other target languages utilizing PCFGs.
First, the program state \rlstinline!PSTATE! consists of a byte array \lstinline!stack! and a pointer to the top of this stack named \lstinline!stackPtr!.
We increment and decrement this stack pointer when stack frames are added and removed, respectively, at function calls and returns.
The type \lstinline!STACK_f! represents the stack frame for the function \lstinline!f! (such a stack frame type must be determined and set up for each function we decompose) and contains its block return address \lstinline!ra!, its parameter \lstinline!p! (functions with multiple parameters have one entry for each parameter), and an address \lstinline!retValLoc! at which we write its return value.
Additionally, it contains the local variables \lstinline!s1!, \lstinline!s3!, and \lstinline!s4! that travel across the blocks in \lstinline!f!.
Note, however, that local variables used only within a single block do not need to go in the stack frame (e.g., \lstinline!t1! and \lstinline!s2!), and the underlying target language (e.g., CUDA for RootPPL) can instead handle them directly.
Lines 13--24 in block 2 in Fig.~\ref{fig:example:c} illustrate the recursive call to \lstinline!f! at line 12 in Fig.~\ref{fig:example:b}.
Here, we allocate a new complete stack frame \lstinline!callsf! and initialize \lstinline!ra!, \lstinline!p!, and \lstinline!retValLoc!.
Allocating the complete stack frame prior to the function call is different from most standard compilers, which most often allocate the part of the stack frame containing local variables at the start of the called function.
This strategy allows for making the allocation size dependent on, e.g., function arguments.
Here, we instead know all stack frame sizes at compile time.
After setting up the stack frame, we increment the stack pointer at lines 21--23 and pass control to the recursive invocation of \lstinline!f! by using \rlstinline!BBLOCK_JUMP! at line 24.
Inversely, we illustrate function return in block 4 on lines 3--7.
First, we set the return value, and second, we decrement the stack pointer.
Finally, we retrieve the return block from the stack frame and pass control to this block at line 7.

\subsection{Function Decomposition Algorithm}\label{sec:decalg}
We now turn to a formal description of the decomposition algorithm.
To avoid going into specifics of the underlying target language, and in particular the call stack handling, we take an abstract view of function bodies and regard them as lists of statements of the form
\begin{equation}\label{eq:stmts}
  \begin{aligned}
    \stmt \Coloneqq \enspace
    \texttt{checkpoint}
    \enspace | \enspace
    \texttt{call}
    \enspace | \enspace
    \texttt{if} \enspace
    [\stmt] \enspace
    [\stmt]
    \enspace | \enspace
    \texttt{other}.
  \end{aligned}
\end{equation}
Here, the $[\stmt]$ syntax indicates a list of $\stmt$s.
Thus, the \texttt{if} construct inductively contains two lists of $\stmt$s---one for each branch.

\begin{figure}[tb]
  \centering
  \begin{subfigure}[c]{0.36\columnwidth}
    \centering
    \begin{lstlisting}[style=normal,style=numbers]
$[$
  other,
  checkpoint,
  other,
  if $[$other$]$ $[$other$]$,
  other,
  if $[$
    other,
    if
      $[$other$]$
      $[$call$]$,
    other,
  $]$ $[$other$]$,
  other
$]$
    \end{lstlisting}
    \caption{The program from Fig.~\ref{fig:example:b} translated to type $[\stmt]$.}
    \label{fig:absexample:a}
  \end{subfigure}%
  \begin{subfigure}[c]{0.64\columnwidth}
    \scriptsize
    \centering
    \lstset{basicstyle=\ttfamily\scriptsize}
      \begin{tikzpicture}[node distance=4mm]

        \node[draw]
          (b1) {%
            \hspace{4pt}
            \begin{lstlisting}[style=normal,style=numbers]
$[$
  other,
  checkpoint $2$
$]$
            \end{lstlisting}
          };
        \node[anchor=south west,inner xsep=0] at (b1.north west) {1};

        \node[%
          draw,right=of b1.north east,anchor=north west,yshift=10pt
          ] (b2) {%
            \hspace{7pt}
            \begin{lstlisting}[style=normal,style=numbers]
$[$
  other
  if
    $[$ other $]$
    $[$ other $]$
  other
  if $[$
    other
    if $[$
      other,
      jump $3$
    $]$ $[$
      call $3$
    $]$
  $]$ $[$
    other,
    jump $4$
  $]$
$]$
            \end{lstlisting}
          };
        \node[anchor=south west,inner xsep=0] at (b2.north west) {2};

        \node[draw,below=20pt of b1.south east,anchor=north east] (b3) {%
            \hspace{4pt}
            \begin{lstlisting}[style=normal,style=numbers]
$[$
  other,
  jump $4$
$]$
            \end{lstlisting}
          };
        \node[anchor=south west,inner xsep=0] at (b3.north west) {3};

        \node[draw,below=20pt of b3.south east,anchor=north east] (b4) {%
            \hspace{4pt}
            \begin{lstlisting}[style=normal,style=numbers]
$[$
  other,
  jump return
$]$
            \end{lstlisting}
          };
        \node[anchor=south west,inner xsep=0] at (b4.north west) {4};

        \node (ra) at ($(b4.south)+(0,-0.5)$) {};
        \node (init) at ($(b1.north)+(0,0.5)$) {};

        \draw[-{Latex[]}]
          ($(b1.south east)+(0,0.67)$)
          -- ++(0.2,0)
          |- ($(b2.north west)-(0,0.8)$);
        \draw[-{Latex[]},gray]
          ($(b2.south west)+(0,3.35)$)
          -- ++(-0.06,0)
          -- ++(-0.06,0)
          |- ($(b3.north east)+(0,-0.8)$);
        \draw[-{Latex[]},gray]
          ($(b3.south east)+(0,0.7)$)
          -- ++(0.2,0)
          |- ($(b4.north east)+(0,-0.8)$);
        \draw[-{Latex[]},gray]
          ($(b2.south west)+(0,1.05)$)
          -- ++(-0.2,0)
          |- ($(b4.north east)+(0,-0.8)$);
        \draw[-{Latex[]},gray] (b4) to (ra);
        \draw[-{Latex[]},gray] (init) to (b1);
      \end{tikzpicture}
      \caption{Decomposition of (a) into $[\tstmt]$ basic blocks.}
    \label{fig:absexample:b}
  \end{subfigure}
  \caption{%
    Illustrating Algorithm~\ref{alg:dec} on the example from Fig.~\ref{fig:example}.
  }
  \label{fig:absexample}
\end{figure}
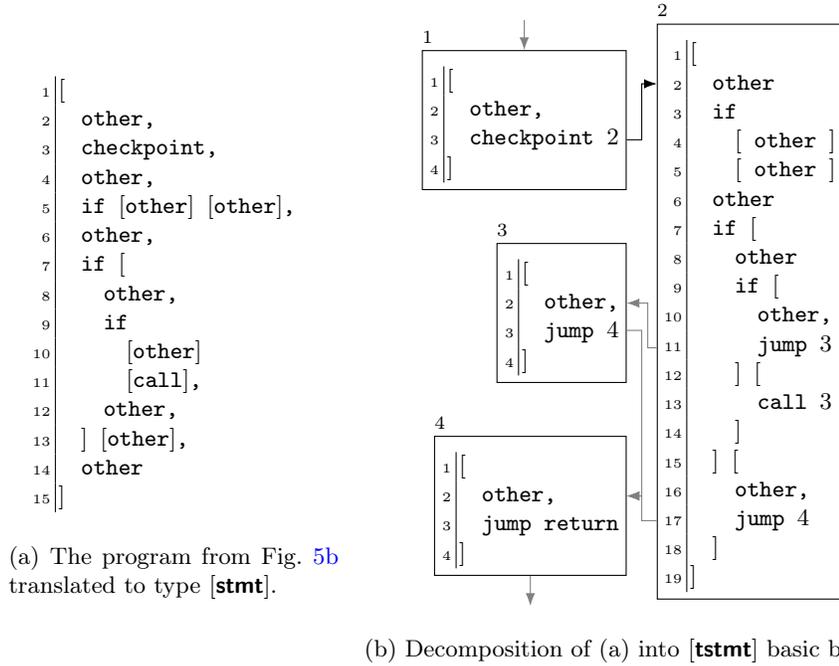

We illustrate the representation $\stmt$ through an example.
Consider the program in Fig.~\ref{fig:example:b} and its mapping to $\stmt$s in Fig.~\ref{fig:absexample:a}.
Due to ANF, we can view the body of \lstinline!f! as a sequence of \clstinline!let! bindings and operations separated by \lstinline!;!, each performing a single operation of some kind (e.g., a checkpoint or a function application).
We map each such operation to a \stmt\ in Fig.~\ref{fig:absexample:a}.
The \clstinline!resample! checkpoint at line 4 in Fig.~\ref{fig:example:b} maps to a \ttt{checkpoint} at line 3 in Fig.~\ref{fig:absexample:a}, and the application of \lstinline!f! at line 12 maps to a \texttt{call} at line 11.
However, other applications, such as \lstinline!geqf! and \lstinline!leqf!, are guaranteed not to encounter any checkpoints.
Therefore, they map to \ttt{other}s, and \emph{not} \ttt{call}s.
The three \clstinline!if!s at lines 6, 9, and 12 map to \ttt{if}s.
Note that we always lift the \clstinline!if! conditions in Fig.~\ref{fig:example:b} to a separate \clstinline!let! as a result of ANF, and they are therefore not part of the \ttt{if} representation in $\stmt$.
We map all remaining operations to \ttt{other}s.

While the illustration above only shows how to map a CorePPL function body to $\stmt$s, the representation is general.
For example, in the CorePPL compiler (Section~\ref{sec:cpplcompile}), the decomposition is performed \emph{after} translation to C, and not at the CorePPL stage.
The reason is that there are no basic blocks in CorePPL.
It is, therefore, more natural to perform this translation closer to RootPPL.

\begin{algorithm}[tbp]
  \newcommand{\s}{\hphantom{|}}
  \caption{%
    A functional-style algorithm for function decomposition into basic blocks.
    We denote tuples with comma-separated expressions within parentheses and sequences with comma-separated items within square brackets.
    We denote type annotation with the : character, the cons operator with :: characters, and sequence concatenation with $\doubleplus$.
    The non-pure function \textsf{newIndex} returns a unique number from $\mathbb{N}$ at every call.
  }\label{alg:dec}
  \ifthesis\vspace{-1mm}\fi
  \hspace{15pt}
  \begin{tabular}{c}
    \begin{lstlisting}[style=alg,numbers=left]
function $\textsc{decompose}$ srcs: $[\stmt] \ra (\mathbb{N} \ra [\tstmt])$ =
  let (block, blocks, _) = $\textsc{rec}$ ($[]$, $\varnothing$, $\ttt{return}$) srcs in
  blocks $\cup$ (newIndex (), block)

function $\textsc{initNext}$ next: $\nextty_+ \ra \nextty$ =
  match next with $\ttt{none}$ $\ra$ newIndex () | _ $\ra$ next

function $\textsc{rec}$ (block, blocks, next) srcs: $\accty \ra [\stmt] \ra \accty$ =
  match srcs with
  | $[] \ra$ match next with
    | $\ttt{none} \ra$ (block, blocks, next)
    | $n$ | $\ttt{return} \ra$ (block $\doubleplus$ $[\ttt{jump} \textnormal{ next}]$, blocks, next)
  | src :: srcs $\ra$ match src with
    | $\ttt{checkpoint}$ | $\ttt{call}$ $\ra$ match srcs with
      | $[] \ra$
        let next = $\textsc{initNext}$ next in
        (block $\doubleplus$ $[$src next$]$, blocks, next)
      | _ ->
        let index = newIndex () in
        let block = block $\doubleplus$ $[$src index$]$ in
        let (nextBlock, blocks, next) = $\textsc{rec}$ ($[]$, blocks, $\textsc{initNext}$ next) srcs in
        (block, blocks $\cup$ (index, nextBlock), next)
    | $\ttt{other} \ra$ $\textsc{rec}$ (block $\doubleplus$ $[\ttt{other}]$, blocks, next) srcs
    | $\ttt{if}$ thn els $\ra$ match srcs with
      | $[] \rightarrow$
        let (thn, thnBlocks, thnNext) = $\textsc{rec}$ ($[]$, blocks, next) thn in
        let (els, elsBlocks, elsNext) = $\textsc{rec}$ ($[]$, thnBlocks, thnNext) els in
        let thn = if next $\neq$ elsNext $\land$ thnNext = $\ttt{none}$
          then thn $\doubleplus$ $[\ttt{jump}$ elsNext$]$ else thn in
        (block $\doubleplus$ $[\ttt{if}$ thn els$]$, elsBlocks, elsNext)
      | _ $\ra$
        let (thn, thnBlocks, thnNext) = $\textsc{rec}$ ($[]$, blocks, $\ttt{none}$) thn in
        let (els, elsBlocks, elsNext) = $\textsc{rec}$ ($[]$, thnBlocks, thnNext) els in
        if elsNext = $\ttt{none}$ then $\textsc{rec}$ (block $\doubleplus$ $[\ttt{if}$ thn els$]$, elsBlocks, next) srcs
        else
          let thn = if thnNext = $\ttt{none}$ then thn $\doubleplus$ $[\ttt{jump}$ elsNext$]$ else thn in
          let (nextBlock, blocks, next) =
            $\textsc{rec}$ ($[]$, elsBlocks, $\textsc{initNext}$ next) srcs in
          (block $\doubleplus$ $[\ttt{if}$ thn els$]$, blocks $\cup$ (elsNext, nextBlock), next)
    \end{lstlisting}
  \end{tabular}
  \ifthesis\vspace{-0.5mm}\fi
\end{algorithm}

We now turn to the full decomposition algorithm over lists of \stmt s, given in Algorithm~\ref{alg:dec}.
The target language representation is a small extension of \stmt, adding transitions between $\mathbb{N}$-indexed basic blocks.
It is given by
\begin{equation}\label{eq:stmtsplus}
  \begin{aligned}
    \tstmt \Coloneqq& \enspace
    \texttt{checkpoint} \enspace \nextty
    \enspace | \enspace
    \texttt{call} \enspace \nextty
    \\ |& \enspace
    \texttt{if} \enspace
    [\tstmt] \enspace
    [\tstmt]
    \enspace | \enspace
    \texttt{jump} \enspace \nextty
    \enspace | \enspace
    \texttt{other}.
  \end{aligned}
\end{equation}
In particular, we annotate \ttt{checkpoint}s and \ttt{call}s with the type \nextty, given by $\nextty \Coloneqq \texttt{return} \enspace | \enspace n$, where $n \in \mathbb{N}$.
For \ttt{checkpoint}s, the \nextty\ indicates which block to jump to after handling the checkpoint, and for \ttt{call}s, it indicates the block to \emph{return to} (e.g., the value set for \rlstinline!ra! in Fig~\ref{fig:example:c}) at the end of the function invocation.
We also include a \ttt{jump} in \tstmt\ for directly jumping to another block (corresponding to \rlstinline!BBLOCK_JUMP! in Fig.~\ref{fig:example:c}).
The \ttt{return} case of \nextty\ indicates that the return address gives the next block for the current function call.
For example, \rlstinline!BBLOCK_JUMP(sf->ra)! is equivalent to \ttt{jump} \ttt{return}.

Fig.~\ref{fig:absexample:b} shows the result of applying Algorithm~\ref{alg:dec} on the $[\stmt]$ in Fig.~\ref{fig:absexample:a}.
Note that the block structure in Fig.~\ref{fig:absexample:b} mirrors that of Fig.~\ref{fig:example:c}.
The entry point in Algorithm~\ref{alg:dec} is the function $\textsc{decompose}$, which accepts a $[\stmt]$ as input, and produces a map from indices to $[\tstmt]$ as output (e.g., Fig~\ref{fig:absexample:b}).
The core of Algorithm~\ref{alg:dec} is the function $\textsc{rec}$, which recursively constructs the basic blocks.
It is called from \textsc{decompose}, and makes use of the function \textsc{initNext}.
The accumulator is the triple \textsf{(block, blocks, next)} of type $\accty \Coloneqq [\stmt] \times (\mathbb{N} \rightarrow [\stmt]) \times \nextty_+$, where \textsf{block} is the current block being constructed, \textsf{blocks} are all blocks constructed so far, and \textsf{next} indicates the action to take at tail position in the current block.
The type $\nextty_+$ is defined as $\nextty_+ \Coloneqq \nextty \enspace | \enspace \texttt{none}$.
When reaching the end of a block, a value \texttt{none} for next means do nothing, a value \texttt{return} indicates that the next block is the return block for the current function invocation, and a natural number $n$ means that the next block has index $n$.

We now walk through the translation of Fig.~\ref{fig:absexample:a} to Fig.~\ref{fig:absexample:b}.
We set the accumulator to $([]$, $\varnothing$, $\ttt{return})$ at line 2 in Algorithm~\ref{alg:dec} just before the initial call to $\textsc{rec}$, indicating that the current block is empty, that we have accumulated no complete blocks so far, and that we must use the return block address when reaching the end of the current block.
In the first call to $\textsc{rec}$, the \ttt{other} at line 2 in Fig.~\ref{fig:absexample:a} triggers the case at line 23 in Algorithm~\ref{alg:dec}, which accumulates the \ttt{other} in the current block.
Next, the \ttt{checkpoint} triggers the case at line 14, followed by line 18, since the \ttt{checkpoint} is not at tail position.
At line 19, we create a new index for the following block.
We then close the current block by tagging the \ttt{checkpoint} with the new index, resulting in block 1 in Fig.~\ref{fig:absexample:b}.
Next, we recursively create the block following the \ttt{checkpoint} at line 21.
Finally, we add the recursively created block with the new index to the map of complete blocks (now also populated by the recursive call) and return the updated accumulator triple at line 22.

The complex part of Algorithm~\ref{alg:dec} involves handling of \texttt{if}s.
In particular, we must handle cases where there are block splits within the branches with care.
In our example, the first \ttt{if} at line 5 in Fig.~\ref{fig:absexample:a} triggers the case at line 31 since it is not in tail position.
To determine whether or not there is at least one split within the branches, we set \textsf{next} to \ttt{none} for the call on line 32.
If a block is split during this call, \textsc{initNext} will be applied on \textsf{next}, and \textsf{thnNext} at line 32 will be a natural number, indicating where the branch jumped to (either through a \ttt{jump}, \ttt{checkpoint}, or \ttt{call}) at tail position.
However, if there is no split in the branch, the resulting \textsf{thnNext} remains \ttt{none}.
There is no split in the first branch of the \ttt{if} at line 5 in Fig.~\ref{fig:absexample:a}, and \ttt{none} is passed to the recursive call at line 33 as well.
Again, there is no split in the second branch, triggering the then case at line 34, and we accumulate the \ttt{if} in the same way as an \ttt{other}.

The \ttt{if}s at lines 7 and 9 in Fig.~\ref{fig:absexample:a} do contain a split due to the \ttt{call} at line 11, resulting in blocks 2, 3, and 4, shown in Fig.~\ref{fig:absexample:b}.
The \textsf{elsNext} is a natural number for these \ttt{if}s, and the else case at line 35 is triggered.
Here, we must take particular care if there is only a split in the second branch of the \ttt{if} and not the first.
In that case, \textsf{thnNext} is \ttt{none}, and unlike the second branch, we do not add a block jump to the end of this branch in the call at line 32.
Therefore, we must instead add it at line 36.
We add the \ttt{jump} at line 11 in block 2 in Fig.~\ref{fig:absexample:b} in this way.
Note that we do not require an equivalent step to the above for the second branch if the split is only in the first branch, since we pass the \textsf{next} from the first branch to the recursive call for the second branch.
After handling the \ttt{if} itself, we recursively create the new block following the \texttt{if} at lines 37--38 (note that we pass the \textsf{next} given as argument to \textsc{rec} here, and use \textsc{initNext} on it to indicate a split has occurred), and give it the index \textsc{elsNext} at line 39.

The case where \ttt{if} is at tail position, at line 25, is handled similarly to the case at line 31.
The difference is that we do \emph{not} pass \ttt{none} to the first branch since there is nothing following the \ttt{if} which we can jump to.
Instead, we directly pass the current \textsf{next} to the first call at line 26.

In the blocks resulting from Algorithm~\ref{alg:dec}, \ttt{call} and \ttt{checkpoint} only occurs in tail-position by construction.
As discussed in Section~\ref{sec:decex}, this is precisely the required property when compiling to PCFGs.

\subsection{CorePPL-to-RootPPL Compiler}\label{sec:cpplcompile}
\begin{figure}[tb]
\centering
\includegraphics[width=0.65\columnwidth]{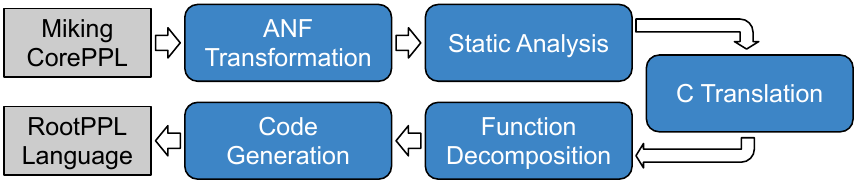}
\caption{%
  The main components of the CorePPL-to-RootPPL compiler.
  Grey blocks are programs, and blue blocks are transformations or analyzes.
}
\label{fig:compiler-overview}
\end{figure}
Fig.~\ref{fig:compiler-overview} gives an overview of the CorePPL-to-RootPPL compiler components.
Besides the techniques described previously, an integral part of the compiler is the C translation step, which translates many of the CorePPL language features to C, including data type definitions and pattern matching.
More precisely, CorePPL records and variants are translated to C structs and tagged unions, respectively, while pattern matching is compiled to C \rlstinline!if! statements.

A simple static analysis phase discovering functions that are guaranteed not to encounter any \lstinline!resample!s is also part of the compiler.
It iterates through all functions and marks a function as containing a resample if it either directly contains a \lstinline!resample! or calls another function containing a \lstinline!resample!.
We do not need to decompose resample-free functions, and invocations can be handled directly by the C++ or CUDA compiler (and we do not need to set up an explicit stack frame).
An example of such a function invocation is the \lstinline!geqf s1 1.! at line 5 in Fig.~\ref{fig:example:b}.
We disallow passing functions as arguments to other functions as it complicates the analysis.
A solution to allow passing functions as arguments is to use static analysis techniques such as 0-CFA \cite{nielson1999principles} instead.

The code generation stage in Fig.~\ref{fig:compiler-overview} adds RootPPL boilerplate code and emits a complete RootPPL program that is provided as input to a C++ or CUDA compiler together with the RootPPL inference engine (see Fig.~\ref{fig:overview}).
The CorePPL compiler implementation is hosted at GitHub~\cite{mikingdppl2021} and consists of approximately 3000 lines of code (a contribution of this paper).
Note that the ANF, static analysis, and C translation steps are quite standard, with no new contributions.

An important detail concerning memory allocation in the compiler is the translation between relative and absolute addresses.
Fig.~\ref{fig:example:c} illustrates this translation.
On line 3 in block 4, we convert the \lstinline!retValLoc! relative pointer to an absolute pointer prior to dereferencing, and at lines 18--20 in block 2, the address of \lstinline!s4! is translated to a relative address with respect to the start of the stack before being assigned to \lstinline!retValLoc!.
This translation is needed because, at checkpoints in RootPPL, resampling copies and moves SMC executions in memory.
Therefore, we cannot use absolute addresses to refer to data on the \rlstinline!PSTATE! stack and must instead use addresses relative to the start of the stack.

\subsection{Compiler Strengths and Limitations}\label{sec:lim}
The main strength of the CorePPL compiler, compared to using other PPL compilers and tools, is the execution time of the compiled programs.
In particular, the compilation from a universal PPL to CUDA is the first of its kind and allows for utilizing GPUs for massively parallel SMC inference.

The compiler does, however, have some limitations.
Most importantly, the lack of standard garbage collectors in C++ and CUDA leads to restrictions for automatic data allocation.
Currently, we support only stack-based allocation, which means that CorePPL programs that allocate and return dynamically sized data structures (e.g., trees or linked lists) from functions are not supported.
Consequently, the current compiler cannot handle probabilistic programs encoding distributions over such data structures (e.g., phylogenetic trees)---the distribution must be over fixed-size data types.
However, as the evaluation in Section~\ref{sec:eval} suggests, practically significant universal probabilistic programs over fixed-sized data types are plentiful.
In general, the compiler supports universal CorePPL programs including both stochastic branching and an unbound number of (stack-allocated) random variables.
Automatic heap-based data allocation is a general challenge when compiling to GPUs and not specific to our approach.
Exploring the use of garbage collectors or other means for automatic memory management on GPUs is an interesting direction for future research.

The compiler also lacks support for some features, which we foresee no substantial technical challenges in implementing in the near future.
In particular, the compiler does not support first-class distributions---we restrict distributions to occur immediately at \clstinline!assume!s (e.g., the Bernoulli distribution in \clstinline!assume (Bernoulli p)! in Fig.~\ref{fig:coreppl:a}).
Another possible feature is to add limited support for nested and higher-order functions.

\section{Evaluation}\label{sec:eval}
This section evaluates RootPPL and the CorePPL-to-RootPPL compiler.
The source code for all experiments is publicly available~\cite{lunden2022compilingartifact}.
We compare RootPPL and CorePPL to state-of-the-art SMC PPL implementations on two models: a constant rate birth-death (CRBD) model from evolutionary biology (Sections~\ref{sec:expcrbd} and~\ref{sec:expocrbd}) and a vector-borne disease model from epidemiology (Section~\ref{sec:expssm}).
Previous work shows that SMC handles these models particularly well~\cite{ronquist2021universal,murray2018delayed}, and they are therefore good candidates for this evaluation.
Comparison with other types of inference algorithms is a challenging problem and beyond the scope of this paper.
For example, comparing SMC with variational inference (VI) is challenging as VI is approximate and SMC is asymptotically exact.

In addition to CorePPL (compiled to RootPPL) and RootPPL (hand-tuned), we implement the models above in a set of state-of-the-art PPLs with SMC inference: Birch~\cite{murray2018automated}, WebPPL~\cite{goodman2014design}, and Pyro~\cite{bingham2019pyro}.
For each PPL, we implement the two models as efficiently as possible, given the available language features.
We compile RootPPL with GCC 7.5.0 for single-core and multicore and with CUDA 11.4 for GPU.
We compile Birch 1.634 with GCC 7.5.0.
We use WebPPL 0.9.15 with Node.js 14.17.6.
We use Pyro 1.7.0 with PyTorch 1.9.0 and CUDA 10.2.
Additionally, we use Numba 0.54.0---a just-in-time (JIT) compiler for Python---to improve the Pyro performance for the Section~\ref{sec:expcrbd} experiment.

To aid the comparison between languages both in the text and in the figures, we use the (S), (M), and (G) symbols suffixed to PPL names to indicate if they run on single-core, multicore, or GPU, respectively.
Despite the CUDA dependency for Pyro, we did not observe any GPU usage during Pyro SMC runs.
In Pyro, SMC is a minor inference algorithm, with variational inference instead being the main focus.
This may explain this lack of GPU support for SMC.
Consequently, we classify SMC in Pyro as (M) and not (G).

We ran all experiments on a machine with a 12-core (24 threads) Intel Xeon Gold 6136 CPU, 64 GB of memory, and an NVIDIA TITAN RTX GPU with 24 GB of memory and 4608 CUDA cores.

\subsection{Experiment: Constant-Rate Birth Death}\label{sec:expcrbd}
In this experiment, we consider the non-trivial CRBD model described in Ronquist et al.~\cite{ronquist2021universal}.
This model encodes the posterior distributions of the rates with which new evolutionary lineages arise (birth rate) and die out (death rate), conditioned on the input of a fixed evolutionary tree (phylogeny).
We use the dated Alcedinidae phylogeny (Kingfisher birds) referenced in Ronquist et al.~\cite{ronquist2021universal}, and introduced in Jetz et al.~\cite{jetz2012global}.
A notable feature of this model is that it contains recursive tree constructions, which are only expressible in universal PPLs.
\ifextended
Appendix~\ref{sec:sourcecode} contains the CorePPL implementation of this model (118 lines of code).
\fi

We measure execution time.
To ensure fairness, we disabled variance-reducing techniques such as delayed sampling~\cite{murray2018delayed} and ESS-triggered resampling in all PPLs where available.
Consequently, all implementations use precisely the same SMC inference algorithm.
We checked this and the implementations' correctness by considering the output normalizing constant estimates in all runs\ifextended\ (see Appendix~\ref{sec:normconst})\fi.
The variance and mean of these estimates were comparable for all PPLs.

\begin{figure}[tb]
  \resizebox{\textwidth}{!}{\input{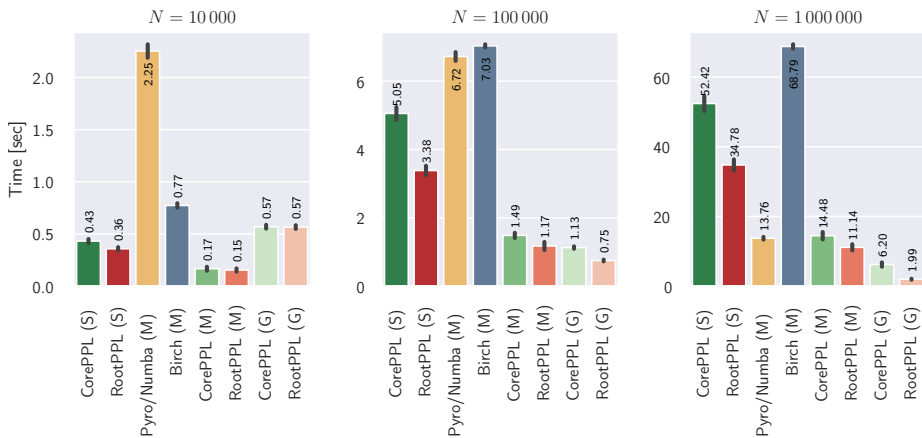}}
  \caption{Execution times for the CRBD experiment, for different numbers of particles $N$. The vertical line at the top of each bar indicates one standard deviation. PPLs with an (S) runs on a single core, (M) on multicore, and (G) on the GPU.}
  \label{fig:expcrbd}
\end{figure}
The results of the experiment are shown in Fig.~\ref{fig:expcrbd} for three different numbers of SMC particles: $10\,000$, $100\,000$, and $1\,000\,000$.
We ran the PPL implementations for 100 iterations (a number determined by available time and hardware) for each number of SMC particles.
The exception to this is WebPPL (S) and Pyro (M), which we ran only for $10\,000$ particles due to excessive execution times.
For $10\,000$ particles, WebPPL~(S) ran for 55 seconds (standard deviation 0.63 seconds), and Pyro (M) for 250 seconds (standard deviation 28 seconds).
We omit WebPPL~(S) and Pyro~(M) from Fig.~\ref{fig:expcrbd}.
Pyro relies heavily upon vectorization through PyTorch, and the expensive operations in the CRBD model are recursive and stochastic tree constructions, which are difficult to vectorize.
This explains the particularly abnormal execution times for Pyro (M).

RootPPL is the best alternative in all categories.
We conjecture that the difference compared to CorePPL is due to hand-tuned details in the RootPPL model.
The RootPPL model uses efficient array encodings of the observed tree, precomputes the recursion order over this tree, and encodes it as an iterative procedure.
CorePPL instead compiles the tree as a tagged union type with pointers to subtrees in each node and traverses it via recursion.
Automatically discovering this transformation from trees to arrays and recursion to iteration is non-trivial and not considered here but could have potential for future work.

To improve the performance of Pyro, we also applied Numba to parallelize the recursive tree construction in the model manually.
The parallelization we apply is more fine-grained than the natural SMC particle parallelism and resulted in an order-of-magnitude performance boost over Pyro (M).
Unlike CorePPL, RootPPL, and Birch, the execution times for Pyro/Numba (M) seems to grow sub-linearly when going from $100\,000$ to $1\,000\,000$ particles, as this only increases mean execution time from 6.72 seconds to 13.76.
We conjecture that this is related to the different type of parallelism introduced with Numba, in combination with its JIT compilation.
Therefore, looking at adding such parallelism to RootPPL and CorePPL is an interesting direction for future work.


\subsection{Experiment: Vector-Borne Disease}\label{sec:expssm}
Next, we consider the vector-borne disease model from Funk et al.~\cite{funk2016comparative}, which is also studied further in Murray et al.~\cite{murray2018delayed}.
This epidemiological model encodes a dengue outbreak in Micronesia and includes the spread of disease between mosquito and human populations.
The inference is over the number of susceptible, exposed, infectious, and recovered (SEIR) individuals in the populations at discrete time steps (days), and the observations are daily numbers of reported new cases at health centers (the data is available in Funk et al.~\cite{funk2016comparative}).
\ifextended
Appendix~\ref{sec:sourcecode} contains the CorePPL implementation of this model (140 lines of code).
\fi

\begin{figure}[tb]
  \resizebox{\textwidth}{!}{\input{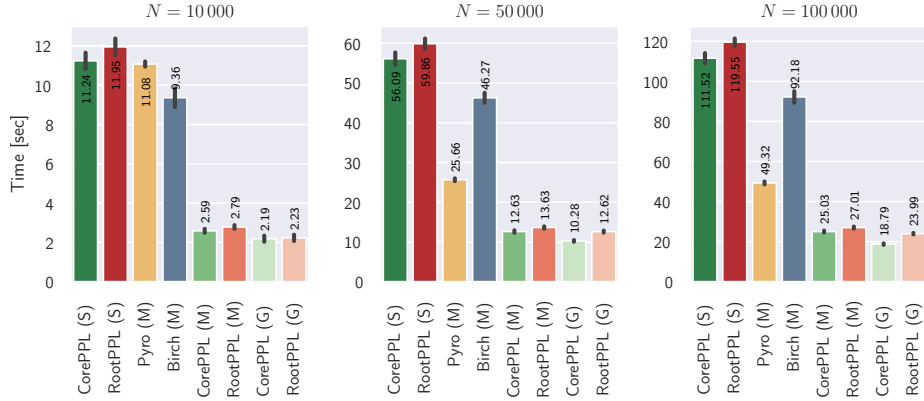}}
  \caption{Execution times for the Vector-Borne Disease experiment, for different numbers of particles $N$. The vertical line at the top of each bar indicates one standard deviation. PPLs with an (S) runs on a single core, (M) on multicore, and (G) on the GPU.}
  \label{fig:expssm}
\end{figure}
The experiment setup is identical to Section~\ref{sec:expcrbd} but with fewer SMC particles due to more demanding computations in the model.
Fig.~\ref{fig:expssm} shows the results.
We omit WebPPL (S) entirely due to high execution times.
However, we include Pyro (M) because the simple non-stochastic control-flow in this model allows much better vectorization than the CRBD model.
The Numba optimization in Section~\ref{sec:expcrbd} relied on the recursive structure of the model.
We exclude Pyro/Numba (M) here, as such an optimization is not possible in this model.

This time, CorePPL is the best option, by a small margin, over RootPPL.
We conjecture that this is due to how RootPPL preallocates memory, which is instead dynamically allocated in CorePPL.
This results in copying slightly more memory during resampling for this model in RootPPL.

The difference between GPU and CPU for CorePPL and RootPPL is not as significant as in Fig.~\ref{fig:expcrbd}.
We conjecture that this is due to the lower numbers of SMC particles used and RootPPL using different implementations for binomial distribution sampling on the CPU and GPU.
The GPU uses a custom, and less efficient version, because the C++ standard library binomial sampling implementation is not available in CUDA.
Because binomial sampling is the most expensive operation in this model, this can improve GPU performance further.

\subsection{Experiment: CRBD with Variance-Reducing Techniques}\label{sec:expocrbd}
In this experiment, we again consider the CRBD model from Section~\ref{sec:expcrbd}, but with delayed sampling and ESS-triggered resampling allowed.
Also, we now consider a different, more challenging phylogeny of Tyrant flycatchers~\cite{ronquist2021universal,jetz2012global}.

\begin{figure}[tb]
  \resizebox{\textwidth}{!}{\input{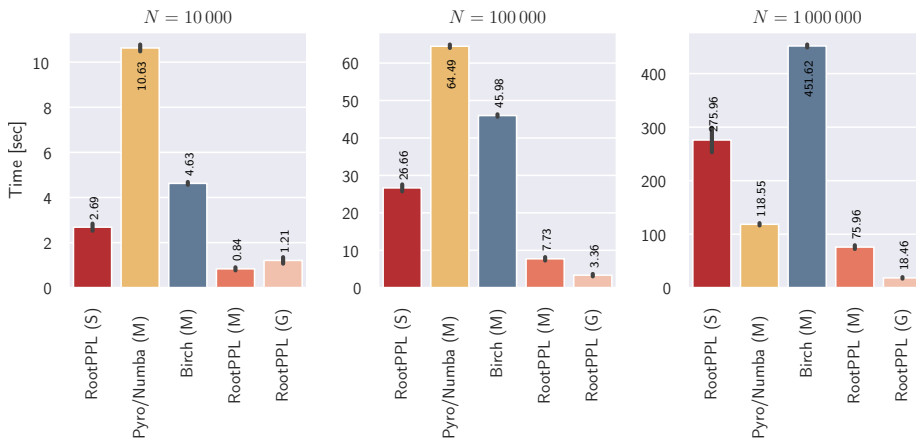}}
  \caption{Execution times for the CRBD experiment with variance-reducing techniques for different numbers of particles $N$. The vertical line at the top of each bar indicates one standard deviation. PPLs with an (S) runs on a single core, (M) on multicore, and (G) on the GPU. Note the 6$\times$ speedup of RootPPL (M) over Birch (M) for $N = 100\,000$.}
  \label{fig:expoptcrbd}
\end{figure}
Fig.~\ref{fig:expoptcrbd} shows the results.
Other than the changes above, the setup is identical to Section~\ref{sec:expcrbd}.
We added static delayed sampling manually to all models to ensure fairness.
Note, however, that automatic and dynamic delayed sampling, as introduced in Murray et al.~\cite{murray2018delayed}, is also natively supported in Birch (but introduces some unfair overhead).
CorePPL is omitted here, as adding efficient delayed sampling to the model is rendered more
difficult by the current lack of support for mutable data structures.
Based on the experiment in Section~\ref{sec:expcrbd}, WebPPL (S) and Pyro (M) are also not considered here.

The results offer no surprise over Fig~\ref{fig:expcrbd}, and RootPPL is again the best alternative.
Note the increased execution times here compared to Fig~\ref{fig:expcrbd} due to the more challenging phylogeny and delayed sampling overhead (which is greatly compensated by increased inference accuracy).


\section{Related Work}\label{sec:relwork}
There are quite a few PPL implementations making use of SMC inference.
Most closely related to the contributions in this paper is Birch \cite{murray2018automated}. Similarly to RootPPL, Birch implements SMC inference, and the target language for compilation is C++.
However, while performance is one of the main goals with Birch, some overhead is inevitably introduced by supporting various quality-of-life C++ features---including automatic heap allocation~\cite{murray2020lazy} and object-oriented features.
RootPPL does not support such features in favor of performance.
Similarly to RootPPL, Birch supports CPU parallelism through the use of OpenMP.
Compilation to GPUs is, however, currently not supported in Birch.

The PCFG concept can also be related to Birch.
In Birch, users write models for SMC inference as a method \lstinline!simulate! which the inference algorithm calls iteratively.
Resampling \emph{only} occurs between calls to this method.
Furthermore, data is passed between calls to \lstinline!simulate! through particle variables stored in an object defined as part of the model (similar to the PCFG state).
We can view PCFG basic blocks as a natural generalization of the Birch \lstinline!simulate! method, conceptually allowing for many \lstinline!simulate! methods with arbitrary control-flow in between them.
In particular, SMC particles can take \emph{different} paths through the PCFG.
As with PCFG blocks, the explicit \lstinline!simulate! function used in Birch can potentially make it more challenging to express models for programmers.
This is not a problem when using our approach of compiling into PCFGs, as we then do the block decomposition automatically.

Besides Birch, parallelism for SMC inference in PPLs is surprisingly absent in previous work.
The predecessor of Birch, LibBi \cite{murray2013bayesian}, is an exception to this and implements highly performant SMC inference through SIMD instructions, OpenMP, and CUDA.
However, in contrast with RootPPL and CorePPL, the LibBi modeling language is not universal.
In other words, LibBi can not express many probabilistic models.

Pyro~\cite{bingham2019pyro} is a PPL mainly focused on stochastic variational inference, supporting MCMC and SMC in addition.
SMC in Pyro is similar to Birch in that models are constructed using an explicit \lstinline!step! function (equivalent to \lstinline!simulate! in Birch).
In general, Pyro supports parallelism through vectorization using PyTorch~\cite{pytorch2021} tensors, which is powerful but also restrictive.
We saw this in Section~\ref{sec:expcrbd}, where we could not use Pyro tensors to parallelize the tree recursion.

Other universal PPLs implementing SMC inference include WebPPL \cite{goodman2014design} and Anglican \cite{wood2014new}.
These languages are embedded in JavaScript, and Clojure, respectively, and implement several inference algorithms (including SMC) through CPS transformations.
The focus is on ease of modeling through functional-style constructs supported by complex runtimes (V8 for JavaScript and the JVM for Clojure) and supporting many different inference algorithms.
Parallelism for SMC is not directly supported, which is different from CorePPL and RootPPL, where the focus is parallelism and performance.

Stan~\cite{carpenter2017stan} and AugurV2~\cite{huang2017compiling} support GPU parallelization of MCMC.
Their modeling languages are, however, more restricted than CorePPL.
Stan supports explicit parallelization of specific functions, and the AugurV2 compiler can compile to MCMC algorithms running partially in parallel on CUDA.
This is quite different from the natural SMC parallelism in CorePPL and RootPPL.

There are also many other probabilistic programming tools, libraries, and languages available, for instance, Gen~\cite{towner2019gen}, Turing~\cite{ge2018turing}, Hakaru~\cite{narayanan2016probabilistic}, and Edward~\cite{tran2016edward}.
Generally, these either focus on assisting users in manually constructing inference algorithms tailored for their specific models or on providing efficient inference for a restricted set of models.

\section{Conclusion}\label{sec:conclude}
This paper introduced the concept of PCFGs and a general method for compiling universal PPLs to PCFGs.
We illustrated these contributions further through the RootPPL implementation and the CorePPL compiler.
This is the first work compiling a universal PPL to GPU with SMC inference.
Furthermore, the evaluation showed that CorePPL and RootPPL can deal with real-world SMC inference problems and outperform the current state-of-the-art with up to 6$\times$ speedups for challenging models (and even more when compared across CPU and GPU).
This gives strong empirical support for the usefulness of the contributions.

Possible improvements upon this work include the exploration of more complex CUDA and C++ runtimes for RootPPL, e.g., runtimes with automatic memory management through garbage collection.
Additionally, high-performance implementations similar to RootPPL for other inference methods (e.g., MCMC) are highly relevant for many probabilistic models---for instance, various models from phylogenetics \cite{ronquist2021universal}.
We leave these topics for future work.

\subsection*{Acknowledgments}
We thank Lawrence Murray for his assistance with Birch; the anonymous reviewers at ESOP for their valuable comments; Gizem Çaylak for her valuable comments and contributions to CorePPL and Miking; Lars Hummelgren, Viktor Palmkvist, and Oscar Eriksson for their valuable comments and contributions to Miking; and finally all other Miking developers for their contributions to Miking.

\clearpage

\bibliographystyle{splncs04}
\bibliography{references}

\ifextended
\clearpage
\appendix

\section{Normalizing Constant Experiment Estimates}\label{sec:normconst}
This section contains normalizing constant estimates produced during the experiments in Section~\ref{sec:eval} and justifies the equivalence and correctness of the implementations in the various PPLs.
Fig.~\ref{fig:normcrbd}, Fig.~\ref{fig:normssm}, and Fig.~\ref{fig:normoptcrbd} show the estimates.

\begin{figure}
  \resizebox{\textwidth}{!}{\input{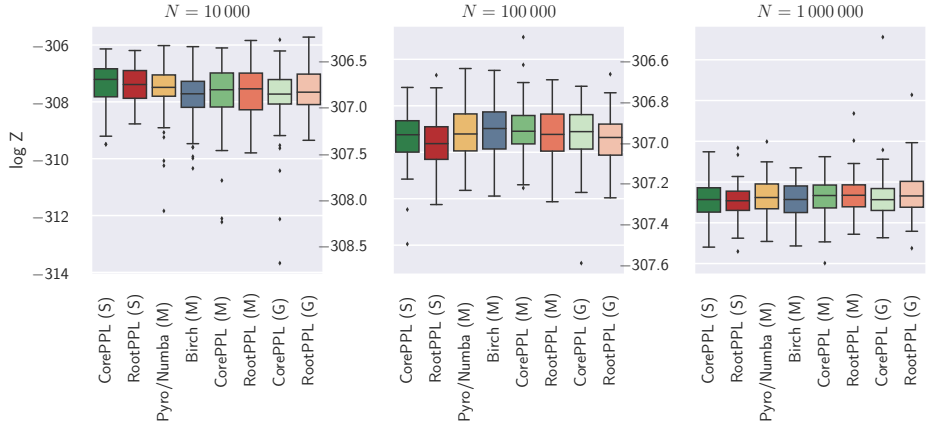}}
  \caption{Box plots of the (log) normalizing constant estimates $Z$ for the CRBD experiment in Section~\ref{sec:expcrbd}. Mirrors the structure of Fig.~\ref{fig:expcrbd}.}
  \label{fig:normcrbd}
\end{figure}

\begin{figure}
  \resizebox{\textwidth}{!}{\input{evaluation/results/output-A-100-experiment-SSM-logz.pgf}}
  \caption{Box plots of the (log) normalizing constant estimates $Z$ for the SSM experiment in Section~\ref{sec:expssm}. Mirrors the structure of Fig.~\ref{fig:expssm}.}
  \label{fig:normssm}
\end{figure}

\begin{figure}
  \resizebox{\textwidth}{!}{\input{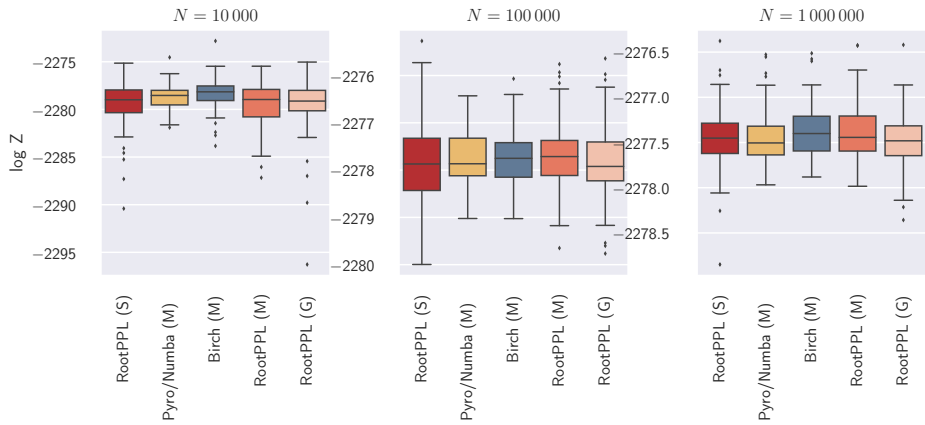}}
  \caption{Box plots of the (log) normalizing constant estimates $Z$ for the CRBD experiment with variance-reducing techniques in Section~\ref{sec:expocrbd}. Mirrors the structure of Fig.~\ref{fig:expoptcrbd}.}
  \label{fig:normoptcrbd}
\end{figure}

\section{CorePPL Experiment Source Code}\label{sec:sourcecode}
This section contains the models used for CorePPL in the experiments presented in Section~\ref{sec:expcrbd} and Section~\ref{sec:expssm}.
These are presented in Listing~\ref{lst:crbd} and Listing~\ref{lst:ssm}, respectively.
\begin{lstlisting}[style=small,style=numbers,language=CorePPL,caption=The CRBD model used for Section~\ref{sec:expcrbd},label=lst:crbd]
------------------------------------------------
-- The Constant-Rate Birth-Death (CRBD) model --
------------------------------------------------

-- The prelude includes a few PPL helper functions
include "pplprelude.mc"

-- The tree.mc file defines the general tree structure
include "tree.mc"

-- The tree-instance.mc file includes the actual tree and the rho constant
include "tree-instance.mc"

mexpr

-- CRBD goes undetected, including iterations. Mutually recursive functions.
recursive
  let iter: Int -> Float -> Float -> Float -> Float -> Float -> Bool =
    lam n: Int.
    lam startTime: Float.
    lam branchLength: Float.
    lam lambda: Float.
    lam mu: Float.
    lam rho: Float.
      if eqi n 0 then
        true
      else
        let eventTime = assume (Uniform (subf startTime branchLength) startTime) in
        if crbdGoesUndetected eventTime lambda mu rho then
          iter (subi n 1) startTime branchLength lambda mu rho
        else
          false

  let crbdGoesUndetected: Float -> Float -> Float -> Float -> Bool =
    lam startTime: Float.
    lam lambda: Float.
    lam mu: Float.
    lam rho: Float.
      let duration = assume (Exponential mu) in
      let cond =
        -- `and` does not use short-circuiting: using `if` as below is more
        -- efficient
        if (gtf duration startTime) then
          (eqBool (assume (Bernoulli rho)) true)
        else false
      in
      if cond then
        false
      else
        let branchLength = if ltf duration startTime then duration else startTime in
        let n = assume (Poisson (mulf lambda branchLength)) in
        iter n startTime branchLength lambda mu rho
in

-- Simulation of branch
recursive
let simBranch: Int -> Float -> Float -> Float -> Float -> Float -> Float =
  lam n: Int.
  lam startTime: Float.
  lam stopTime: Float.
  lam lambda: Float.
  lam mu: Float.
  lam rho: Float.
    if eqi n 0 then 0.
    else
      let currentTime = assume (Uniform stopTime startTime) in
      if crbdGoesUndetected currentTime lambda mu rho then
        let v = simBranch (subi n 1) startTime stopTime lambda mu rho in
        addf v (log 2.)
      else
        negf inf
in

-- Simulating along the tree structure
recursive
let simTree: Tree -> Tree -> Float -> Float -> Float -> () =
  lam tree: Tree.
  lam parent: Tree.
  lam lambda: Float.
  lam mu: Float.
  lam rho: Float.
    let lnProb1 = mulf (negf mu) (subf (getAge parent) (getAge tree)) in
    let lnProb2 = match tree with Node _ then log lambda else log rho in

    let startTime = getAge parent in
    let stopTime = getAge tree in
    let n = assume (Poisson (mulf lambda (subf startTime stopTime))) in
    let lnProb3 = simBranch n startTime stopTime lambda mu rho in

    weight (addf lnProb1 (addf lnProb2 lnProb3));
    resample;

    match tree with Node { left = left, right = right } then
      simTree left tree lambda mu rho;
      simTree right tree lambda mu rho
    else ()
in

-- Priors
let lambda = assume (Gamma 1.0 1.0) in
let mu = assume (Gamma 1.0 0.5) in

-- Adjust for normalizing constant
let numLeaves = countLeaves tree in
let corrFactor =
  subf (mulf (subf (int2float numLeaves) 1.) (log 2.)) (lnFactorial numLeaves) in
weight corrFactor;
resample;

-- Start of the simulation along the two branches
(match tree with Node { left = left, right = right } then
  simTree left tree lambda mu rho;
  simTree right tree lambda mu rho
else ());

lambda

-- Returns the posterior for the lambda
\end{lstlisting}
\begin{lstlisting}[style=small,style=numbers,language=CorePPL,caption=The SSM model used for Section~\ref{sec:expssm},label=lst:ssm]
----------------------------------------------------------------------------
-- The SEIR model from https://docs.birch.sh/examples/VectorBorneDisease/ --
----------------------------------------------------------------------------

-- Include pow and exp
include "math.mc"

-- Include data
include "data.mc"

mexpr

-- Human parameters
let hNu: Float = 0. in
let hMu: Float = 1. in
let hLambda: Float = assume (Beta 1. 1.) in
let hDelta: Float =
  assume (Beta (addf 1. (divf 2. 4.4)) (subf 3. (divf 2. 4.4))) in
let hGamma: Float =
  assume (Beta (addf 1. (divf 2. 4.5)) (subf 3. (divf 2. 4.5))) in

-- Mosquito parameters
let mNu: Float = divf 1. 7. in
let mMu: Float = divf 6. 7. in
let mLambda: Float = assume (Beta 1. 1.) in
let mDelta: Float =
  assume (Beta (addf 1. (divf 2. 6.5)) (subf 3. (divf 2. 6.5))) in
let mGamma: Float = 0. in

-- Other parameters
let rho: Float = assume (Beta 1. 1.) in
let z: Int = 0 in

-- Human SEIR component
let n: Int = 7370 in
let hI: Int = addi 1 (assume (Poisson 5.0)) in
let hE: Int = assume (Poisson 5.0) in
let hR: Int =
  floorfi (assume (Uniform 0. (int2float (addi 1 (subi (subi n hI) hE))))) in
let hS: Int = subi (subi (subi n hE) hI) hR in

-- Human initial deltas
let hDeltaS: Int = 0 in
let hDeltaE: Int = hE in
let hDeltaI: Int = hI in
let hDeltaR: Int = 0 in

-- Mosquito SEIR component
let u: Float = assume (Uniform (negf 1.) 2.) in
let mS: Int = floorfi (mulf (int2float n) (pow 10. u)) in
let mE: Int = 0 in
let mI: Int = 0 in
let mR: Int = 0 in

-- Mosquito initial deltas
let mDeltaS: Int = 0 in
let mDeltaE: Int = 0 in
let mDeltaI: Int = 0 in
let mDeltaR: Int = 0 in

-- Conditioning function
let condition: Int -> Int -> Int -> Int =
  lam t: Int. lam zP: Int. lam hDeltaI: Int.
    let z: Int = addi zP hDeltaI in
    let y: Int = get ys t in
    let z: Int = if neqi (negi 1) y then
      observe y (Binomial z rho); 0
      else z in
    resample;
    z
in

-- Initial conditioning
let z = condition 0 z hDeltaI in

-- Simulation function
recursive let simulate:
  Int -> Int -> Int -> Int -> Int -> Int -> Int -> Int -> Int -> Int -> ()
  =
  lam t: Int.
  lam hSP: Int.
  lam hEP: Int.
  lam hIP: Int.
  lam hRP: Int.
  lam mSP: Int.
  lam mEP: Int.
  lam mIP: Int.
  lam mRP: Int.
  lam zP: Int.

    -- Humans
    let hN: Int = addi (addi (addi hSP hEP) hIP) hRP in
    let hTau: Int =
      assume (Binomial hSP (subf 1. (exp (negf (divf
                                                (int2float mIP)
                                                (int2float hN)))))) in
    let hDeltaE: Int = assume (Binomial hTau hLambda) in
    let hDeltaI: Int = assume (Binomial hEP hDelta) in
    let hDeltaR: Int = assume (Binomial hIP hGamma) in
    let hS: Int = subi hSP hDeltaE in
    let hE: Int = subi (addi hEP hDeltaE) hDeltaI in
    let hI: Int = subi (addi hIP hDeltaI) hDeltaR in
    let hR: Int = addi hRP hDeltaR in

    -- Mosquitos
    let mTau: Int =
      assume (Binomial mSP (subf 1. (exp (negf (divf
                                               (int2float hIP)
                                               (int2float hN)))))) in
    let mN: Int  = addi (addi (addi mSP mEP) mIP) mRP in
    let mDeltaE: Int = assume (Binomial mTau mLambda) in
    let mDeltaI: Int = assume (Binomial mEP mDelta) in
    let mDeltaR: Int = assume (Binomial mIP mGamma) in
    let mS: Int = subi mSP mDeltaE in
    let mE: Int = subi (addi mEP mDeltaE) mDeltaI in
    let mI: Int = subi (addi mIP mDeltaI) mDeltaR in
    let mR: Int = addi mRP mDeltaR in

    let mS: Int = assume (Binomial mS mMu) in
    let mE: Int = assume (Binomial mE mMu) in
    let mI: Int = assume (Binomial mI mMu) in
    let mR: Int = assume (Binomial mR mMu) in

    let mDeltaS: Int = assume (Binomial mN mNu) in
    let mS: Int = addi mS mDeltaS in

    let z: Int = condition t zP hDeltaI in

    -- Recurse
    let tNext: Int = addi t 1 in
    if eqi (length ys) t then
      -- We do not return anything here, but simply run the model for
      -- estimating the normalizing constant.
      ()
    else
      simulate tNext hS hE hI hR mS mE mI mR z
in

-- Initiate recursion
simulate 1 hS hE hI hR mS mE mI mR z
\end{lstlisting}
\fi

\end{document}